\documentclass[11pt]{article}
\usepackage{color}
\usepackage{amssymb}
\usepackage{amsmath}
\usepackage{amsfonts}
\usepackage{pifont}
\usepackage{amsmath,amsfonts,amssymb,theorem,graphicx,listings,txfonts}
\usepackage{graphics}
\usepackage[linesnumbered,ruled]{algorithm2e}
\usepackage{flafter}
\usepackage{indentfirst}
\usepackage{epsfig}
\usepackage{mathrsfs}
\usepackage{subfigure}
\usepackage{cite}
\newtheorem{theorem}{Theorem}[section]

\newtheorem{proposition}[theorem]{Proposition}
\newtheorem{corollary}[theorem]{Corollary}
\newtheorem{definition}[theorem]{Definition}
\newtheorem{example}[theorem]{Example}

\renewcommand{\arraystretch}{1}

\title{Compression of dynamic fuzzy relation
information systems}

\author
{Mingjie Cai$^{a}$ \hspace{1cm} Qingguo Li$^{a}$
\thanks{Corresponding author.\quad Tel./fax: +86 731 88822855,
liqingguoli@aliyun.com
\newline\mbox{}\hspace{0.55cm}
E-mail address: mjcaiphd@gmail.com(Mingjie Cai). } \hspace{1cm}
Guangming Lang$^{b}$\\
\small {$^{a}$ College of Mathematics and Econometrics, Hunan University}\\
\small {Changsha, Hunan 410082, P.R. China}\\
\small {$^{b}$ School of Mathematics and Computer Science, Changsha University of Science and Technology}\\
\small {Changsha, Hunan 410082, P.R. China} }
\date{}

\begin{document}
\maketitle \baselineskip=17pt
\begin{center}
\begin{quote}
{{\bf Abstract.} The notion of homomorphism, as an important tool for studying the relationship between two information systems, has attracted a great deal of attention in recent years. However, in the existing studies, the authors tend to pay their attention to static information systems. In the present paper, we aim to study homomorphisms between fuzzy  relation information systems(FRISs) in dynamic environments. The term 'dynamic' refers to the fact that the involved information systems need to be updated with time, due to inflow of new information. More specifically, we firstly examine properties of consistent functions and construct homomorphisms between FRISs. Such a notion provides a novel approach of reductions in FRISs. Then, we develop
incremental mechanisms for compressing dynamic FRIS. Lastly, several illustrative examples are
employed to demonstrate that constructing homomorphisms between
dynamic FRISs can be simplified
significantly with the proposed algorithms.

\\}
\end{quote}
\end{center}
\renewcommand{\thesection}{\arabic{section}}

\section{Introduction}
Rough set theory, originated by Pawlak, has become a well-established mechanism for uncertainty management  in a wide variety of applications related to artificial intelligence.

The notion of attribute reduction has become one of the most
important issues in the study of rough set theory. Almost since the  inception of rough set theory, many different approaches for solving the issue of attribute reduction have emerged. Homomorphism between information systems, among others, have gained more and more attention in the recent years.
Such a novel notion of homomorphism was initially introduced in \cite{Grzymala-Busse2}. Some of the basic properties were also investigated. Then Li et al.\cite{Li2} explored invariant characters of information systems
under some homomorphisms. Afterwards, many
scholars\cite{Gong1,Wang1,Wang2,Zhu2,Zhu3} discussed the
relationship between information systems by means of homomorphisms.
Wang et al, among  others, \cite{Wang2} investigated homomorphisms
between FRIS. Zhu et al.\cite{Zhu3}
discussed more properties of consistent functions and fuzzy relation
mappings. As illustrated in the
literatures\cite{Grzymala-Busse2,Li2,Gong1,Wang1,Wang2,Zhu2,Zhu3},
Considerable emphasis is given to the computation of the partition of the universe
when constructing homomorphisms between information systems. But
also, the existing studies focus more attention on studying properties of homomorphisms
and few efforts have been made on improving algorithms of
constructed homomorphism. Then, how to
introduce an algorithm of deriving the partition with low
computational complexity becomes a necessity. In addition, FRIS vary with time due to dynamic characteristics of data
collection, and the non-incremental approach to compressing dynamic
FRIS is often very costly or even
intractable. Although there exist some studies on dynamic
information
systems\cite{Wang3,Huang1,Dey1,Zhang1,Liu2,Liu3,Li1,Chen1}, little
attention has been paid to compress dynamic FRIS. Therefore, it is interesting to apply an
incremental updating scheme to maintain the compression dynamically
and avoid unnecessary computations by utilizing the compression of
the original FRIS.

The purpose of this paper is to further study data compression of
FRIS. First, we investigate more
properties of consistent functions and present an algorithm of
compressing FRIS. More concretely, on
the basis of the maximum-consistent function, we construct a
partition of the universe and homomorphisms between FRIS. It is shown that the set of all
consistent functions with respect to a fuzzy relation is a complete
lattice. Subsequently, FRIS can be
compressed into relatively small ones by means of homomorphisms.
Second, we compress dynamic FRIS by
utilizing the precious compressions of the original information
systems. There are five types of dynamic FRIS: immigration and emigration of fuzzy relations and objects,
respectively, variation of fuzzy relation values. We present the
characterizations of five types of dynamic FRIS and employ several examples to illustrate the
process of compressing the dynamic FRIS. Using the proposed approach, the computational complexity
of computing reductions of FRIS
can be reduced greatly by avoiding unnecessary computations.

The rest of this paper is organized as follows: Section 2 briefly
reviews the basic concepts of lattices and FRIS. Section 3 study some properties of
consistent functions and the method of constructing homomorphisms between two FRISs. Section 4 is devoted to compressing
five types of dynamic FRIS. Section 5 are conclusions.

\section{Preliminaries}

In this section, we briefly review the concepts of lattice, FRIS, consistent function, homomorphism and reduction.

\begin{definition}\cite{Birkhoff}\label{def:lattice}
 Let $(L, \preceq)$ be a   non-empty ordered set,

 $(1)$ if $a\vee b$ and  $a\wedge b$  exist for all $a,b\in
 L$, then L is called a lattice, where $a\vee b(resp., a\wedge b)$ denotes the least(resp., the largest lower bound) upper bound of a and b with respect to $\preceq$;

$(2)$ if $\bigvee S$ and  $\bigwedge S$  exist for all
 $S\subseteq L$, then L is called a complete lattice, where $\bigvee S(resp., \bigwedge S)$ denotes the least(resp., the largest lower bound) upper bound of S with respect to $\preceq$;
 and $\bigwedge S=inf(S)$;

$(3)$ if $L$ is a lattice satisfying the distributive law, i.e., $\forall a, b, c \in L$, $a\wedge
(b\vee c)=(a\wedge b)\vee (a\wedge c)$, then
$L$ is said to be distributive;

$(4)$ if $L$ satisfies the modular law, i.e., $\forall a, b, c \in L$, $c\preceq a\Longrightarrow
a\wedge (b\vee c)=(a\wedge b)\vee c$, then $L$
is said to be modular.
\end{definition}

Obviously, every finite lattice is complete according to Definition \ref{def:lattice}. In what follows, we present the concept of FRIS.

\begin{definition}\cite{Wang2}\label{def:FRIS}
Let $U$ be a finite universe, $\mathscr{R}$ a family of fuzzy binary
relations on $U$, where $U=\{x_{1},x_{2},...,x_{n}\}$,
$\mathscr{R}=\{R_{1},R_{2},...,R_{m}\}$ and $R_{i}$ $(1\leq i\leq
m)$ is a fuzzy binary relation on $U$, or in other words, $R_i$ is a mapping from $U\times U$ to $[0,1]$. Then the pair
$(U,\mathscr{R})$ is referred to as a FRIS.
\end{definition}

If the mapping $R\in \mathscr{R}$ takes values from the set $\{0,1\}$,
then $(U,\mathscr{R})$ is a crisp information system\label{def:relationIS}. For
the sake of convenience, we denote all fuzzy binary relations on $U$ by
$\mathscr{F}(U\times U)$.


\begin{definition}\cite{Wang2}\label{def:consistent}
Let $U_{1}$ and $U_{2}$ be two finite universes, $f$ a mapping from
$U_{1}$ to $U_{2}$, $R\in  \mathscr{F}(U_{1}\times U_{1})$, and
$[x]_{f}=\{y\in U_{1}|f(x)=f(y)\}$. $\forall x,y\in U_{1}$, if
$R(u,v)=R(s,t)$ holds for any two pairs $(u,v),(s,t)\in
[x]_{f}\times[y]_{f}$, then $f$ is said to be consistent with
respect to $R$.
\end{definition}

\begin{definition}\cite{Wang2}\label{def:f-inducedFRIS}
Let $U_{1}$ and $U_{2}$ be two finite universes, $f$ be a mapping from $U_{1}$ to $U_{2}$ and $\mathscr{R}=\{R_1, R_2, \cdots, R_n\}$ a family of fuzzy binary relations on $U_{1}$, \underline{$f(\mathscr{R})=\{f(R_1),f(R_2), \cdots, f(R_n)\}$ $\footnote{defination not clear}$}. Then
the pair $(U_{2},f(\mathscr{R}))$ is referred to as an f-induced FRIS of $(U_{1},\mathscr{R})$.
\end{definition}

\begin{definition}\cite{Wang2}\label{def:homomorphism}
Let $(U_{1},\mathscr{R})$ be a FRIS and $(U_{2},f(\mathscr{R}))$ a f-induced FRIS of $(U_{1},\mathscr{R})$. If $\forall R_i \in \mathscr{R}$, f is consistent with respect to $R_i$ on $U_{1}$, then f is referred to as homomorphism from $(U_{1},\mathscr{R})$ to $(U_{2},f(\mathscr{R}))$.
\end{definition}

For the sake of convenience, We always call $(U_{1},\mathscr{R})$ in the above two definitions an original system, and call  $(U_{2},f(\mathscr{R}))$ an image system. Wang et al. discussed the
relationship between FRIS and proved
that reductions of the original and image systems were equivalent to each other in sense of homomorphisms.

\begin{definition}\cite{Wang2}\label{def:reduction}
Let $(U_{1},\mathscr{R})$ be a FRIS, A subset $\textbf{P}\subseteq \mathscr{R}$ is called a reduction of $\mathscr{R}$ if \textbf{P} satisfies the following conditions: \\
(1) $\cap\textbf{P}=\cap\mathscr{R}$;\\
(2) $\forall R_i \in \textbf{P}, \cap\textbf{P}\subset \cap (\textbf{P}-R_i)$.
\end{definition}

\begin{theorem}\cite{Wang2}
Let $(U_{1},\mathscr{R})$ be a FRIS, $(U_{2},f(\mathscr{R}))$ an f-induced FRIS of $(U_{1},\mathscr{R})$ and f be a homomorphism from $(U_{1},\mathscr{R})$ to $(U_{2},f(\mathscr{R}))$ and $\textbf{P}\subseteq\mathscr{R}$. Then $\textbf{P}$ is a reduction of $\mathscr{R}$ if and only if $f(\textbf{P})$ is a reduction of $f(\mathscr{R})$.
\end{theorem}

The consistent function is a
homomorphism between FRISs if it is a
\underline{surjection}. Additionally, the consistent functions provide an
approach to \underline{study relation reduction of FRIS}.

Actually, there are several consistent functions with respect to the
same fuzzy relation, which is illustrated by the following example.

\begin{example}\label{exam:U1R1}
Table 1 depicts the fuzzy relation $R_{1}$ on $U_{1}$, where
$U_{1}=\{x_{1},x_{2},...,x_{8}\}$.
\begin{table}[h!]
\caption{An fuzzy relation information system $(U_{1},R_{1})$}
\tabcolsep0.26in
\label{tab:U1R1}
\begin{tabular}{c c c c c c c c c c c c c c c c c c c c c c c c c c c}
\hline $R_{1}$  & $x_{1}$ &$x_{2}$& $x_{3}$& $x_{4}$& $x_{5}$
&$x_{6}$& $x_{7}$& $x_{8}$\\
\hline
$x_{1}$ & $0.7$& $0.4$ &$0.7$& $0.5$ & $0.7$& $0.4$ &$0.7$& $0.5$\\
$x_{2}$ & $0.7$& $0.3$ &$0.7$& $0.8$ & $0.7$& $0.3$ &$0.7$& $0.8$\\
$x_{3}$ & $0.7$& $0.4$ &$0.7$& $0.5$ & $0.7$& $0.4$ &$0.7$& $0.5$\\
$x_{4}$ & $0.6$& $0.3$ &$0.6$& $0.8$ & $0.6$& $0.3$ &$0.6$& $0.8$\\
$x_{5}$ & $0.7$& $0.4$ &$0.7$& $0.5$ & $0.7$& $0.4$ &$0.7$& $0.5$\\
$x_{6}$ & $0.7$& $0.3$ &$0.7$& $0.8$ & $0.7$& $0.3$ &$0.7$& $0.8$\\
$x_{7}$ & $0.7$& $0.4$ &$0.7$& $0.5$ & $0.7$& $0.4$ &$0.7$& $0.5$\\
$x_{8}$ & $0.6$& $0.3$ &$0.6$& $0.8$ & $0.6$& $0.3$ &$0.6$& $0.8$\\
\hline
\end{tabular}
\end{table}

We define two consistent functions with respect to $R_{1}$
as follows:

$(1)$ $f_{1}$ is a mapping from $U_{1}$ to $U_{2}$, where $U_{2}=\{y_{1}, y_{2}, y_{3}\}$,

\begin{table}[htb]
\tabcolsep0.8in
\label{tab:map1}
\begin{tabular}{c c c}
$x_{1},x_{3},x_{5},x_{7}$ & $x_{2},x_{6}$ & $x_{4},x_{8}$\\
\hline
$y_{1}$ & $y_{2}$ & $y_{3}$ \\
\end{tabular}
\end{table}

Then $U_1$ is compressed into $U_2$ by $f_1$, $f_1$ is consistent with respect to $R_1$.
\begin{table}[htb]
\caption{An fuzzy relation information system $(U_2,f_1(R_1))$}
\tabcolsep0.7in
\label{tab:U2}
\begin{tabular}{c c c c}
\hline
$U_{2}$ & $y_{1}$ & $y_{2}$ & $y_{3}$\\
\hline
$y_{1}$ & 0.7 & 0.4 & 0.5\\
$y_{2}$ & 0.7 & 0.4 & 0.5\\
$y_{3}$ & 0.7 & 0.4 & 0.5\\
\hline
\end{tabular}
\end{table}

$(2)$ $f_{2}$ is  a mapping from $U_{1}$ to $U_{3}=\{z_{1}, z_{2}, z_{3}, z_{4}\}$,

\begin{table}[htb]
\tabcolsep0.6in
\label{tab:map2}
\begin{tabular}{c c c c}
$x_{1},x_{3}$ & $x_{2},x_{6}$ & $x_{4},x_{8}$ & $x_{5},x_{7}$\\
\hline
$z_{1}$ & $z_{2}$ & $z_{3}$ & $z_{4}$\\
\end{tabular}
\end{table}
Then $U_1$ is compressed into $U_3$, $f_2$ is consistent with respect to $R_1$.
\begin{table}[htb]
\caption{An fuzzy relation information system $(U_3,f_2(R_1))$}
\tabcolsep0.54in
\label{tab:U3}
\begin{tabular}{c c c c c}
\hline
$U_{3}$ & $z_{1}$ & $z_{2}$ & $z_{3}$ & $z_{4}$\\
\hline
$z_{1}$ & 0.7 & 0.4 & 0.5 & 0.7\\
$z_{2}$ & 0.7 & 0.4 & 0.5 & 0.7\\
$z_{3}$ & 0.7 & 0.4 & 0.5 & 0.6\\
$z_{4}$ & 0.7 & 0.4 & 0.5 & 0.7\\
\hline
\end{tabular}
\end{table}

\end{example}

Obviously, $f_1$ and $f_2$ are both homomorphisms with respect to $(U_1,R_1)$. Therefore $(U_{1},R_{1})$ can be compressed into different
information systems by means of $f_{1}$ and $f_{2}$, and the image systems i.e., $(U_2, f_1(R_1))$ and $(U_3, f_2(R_1))$ have universes of different cardinality. It is the
interesting to investigate the relationship between two consistent
functions.


%
%
%
%

\section{Compressing FRIS under homomorphisms }

Wang et al. introduced the concept of consistent
functions for constructing homomorphisms between FRISs, but  the constructive methods of consistent functions are not explicitly given. Actually, for a FRIS there may exist a family of consistent
functions. Each consistent function
is a mapping from the original system to the image system. The purpose of compressing original system is to reduce the number of objects in universe, such compressing will improve of computing. Since an image system containing less objects is more easier to deal with in the process of computing than an image system containing more objects. It is necessary to  discuss the relationships between consistent functions, which in turn can help us to obtain the smallest-scale image system under homomorphisms between FRISs.

Firstly, we propose the concept of a partition induced by a fuzzy relation.

\begin{definition}\label{def:partition}
Let $U_{1}$ be a finite universe, the fuzzy relation $R$ a mapping from $U_{1}\times U_{1}$ to $[0,  1]$. Denote
$$[x]_{R}=\{y \in U_1 \mid \forall z\in U_{1},R(x,z)=R(y,z)\}.$$
Then we call $C_{R}=\{[x]_{R}|x\in U_{1}\}$ a partition the fact that $C_R$ is a partition can be easily verified of $U_{1}$ with respect to $R$, and call $[x]_{R}$ an equivalence class of $x$ on $R$.
\end{definition}

\begin{example}(Continued from Example \ref{exam:U1R1})
$U_1=\{x_1,x_2,x_3,x_4,x_5,x_6,x_7,x_8\}$, $R_1$ is the fuzzy relation from $U_1 \times U_1$ to [0,1], as shown in Table \ref{tab:U1R1}. Then we have
$[x_1]_{R_1}=[x_3]_{R_1}=[x_5]_{R_1}=[x_7]_{R_1}$, $[x_2]_{R_1}=[x_6]_{R_1}$, $[x_4]_{R_1}=[x_8]_{R_1}$.
The partition of $U_1$ with respect to $R_1$ is described as follows:
$C_{R_1}=\{\{x_1,x_3,x_5,x_7\},\{x_2,x_6\},\{x_4,x_8\}\}$
\end{example}

Consequently, we show the
relationship between the partitions of the universe and consistent functions.

\begin{theorem}\label{theo:partitionconsistent}
Let $U_{1}$ and $U_{2}$ be two finite universes, \underline{$f$ be a mapping from
$U_{1}$ to $U_{2}$}, $R$ be a fuzzy relation on $U_{1}$. \underline{If $f$ is a consistent function with
respect to $R$}, then $[x]_{f}\subseteq [x]_{R}$ holds for any $x\in
U_{1}$.
\end{theorem}

\noindent{\bf Proof.} Taking arbitrarily $x,z\in U_{1}$, we have
$R(x,z)=R(x_{0},z)$ for any $x_{0}\in [x]_{f}$ since $f$ is a
consistent function with respect to $R$. By Definition \ref{def:partition}, we
have $[x]_{R}=\{y|R(x,z)=R(y,z),z\in U_{1}\}$, which implies
$[x]_{f}\subseteq [x]_{R}$. Therefore, $[x]_{f}\subseteq [x]_{R}$ holds
for any $x\in U_{1}$. $\Box$

The converse of Theorem \ref{theo:partitionconsistent} does not hold in the general case, which can be seen from Example \ref{exam:U1R1}. Now, we present the concepts of $\leq $,
$\vee$ and $\wedge$ for studying the relationship between consistent
functions.

\begin{definition}\label{def:consistentrelation}
Let $U_{1}$ and $U_{2}$ be two finite universes, $I$ be
an indexed set, $f_{i}$ $(i \in I)$ be a
mapping from $U_{1}$ to $U_{2}$, $R$ be a fuzzy relation on $U_1$, $\mathscr{F}_{R}=\{f_{i}|i \in
I\}$ the set of all consistent functions with respect to $R$. Define two binary relations $=$ and $\leq$ in the following manner:

%
%
%
$(1)$ \underline{$f_{1} \approx f_{2} \Leftrightarrow \forall x\in U_1, [x]_{f_{1}} = [x]_{f_{2}}$},

$(2)$ $f_{1}\leq f_{2} \Leftrightarrow \forall x\in U_1, [x]_{f_{1}} \leq [x]_{f_{2}}$.
\end{definition}

It is important to note that the binary relation $\approx$ does not coincide with the usual identity relation $=$. It does not directly depend on the value of any consistent function, but rely heavily on the fact that which two elements in $U_1$ will have the same image under the consistent function. An easy verification shows that $\leq$ is an order on $\mathscr{F}_R$, it can be shown that $[x]_{f_{1}\vee f_{2}}=[x]_{f_{1}}\cup [x]_{f_{2}}$ and
$[x]_{f_{1}\wedge f_{2}}=[x]_{f_{1}}\cap [x]_{f_{2}}$.

\begin{theorem}
Let $U_{1}$ and $U_{2}$ be two finite universes, $R$ be a fuzzy relation on $U_1$, $f_{i}(i=1,2,3,4)$ be a mapping from $U_{1}$ to $U_{2}$, and \underline{
$f_{1},f_{2}\in\mathscr{F}_{R}$.} \footnote{$f_{1}$, $f_{2}$ is consistent}

$(1)$ If $[x]_{f_{3}}=[x]_{f_{1}\vee f_{2}}$, then $f_{3}=f_{1}\vee
f_{2}$;

$(2)$ If $[x]_{f_{4}}=[x]_{f_{1}\wedge f_{2}}$, then
$f_{4}=f_{1}\wedge f_{2}$;

$(3)$ If $f_3=f_{1}\vee
f_{2}, f_4=f_{1}\wedge
f_{1}$ $f_{2}$, then $f_{3}$ and $f_{4}$ are consistent functions with respect to
$R$.
\end{theorem}

\noindent{\bf Proof.} (1) and (2) follow immediately from Definition \ref{def:consistentrelation}.

%
(3) According to Definition \ref{def:consistent}, it suffices to show that for all $(u, v)$, $(s, t) \in [x]_{f_3} \times [y]_{f_3}$, $R(u, v)=R(s, t)$ holds. Indeed, for $u \in [x]_{f_3}$, we have $u \in [x]_{f_1}$ or $u \in [x]_{f_2}$, considering the fact that both $f_1$ and $f_2$ are consistent functions with respect to $R$, we conclude $R(x, v)=R(s, t)$ by using Definition \ref{def:consistent}. Similarly, we can show that $R(x, v)=R(x, y)$, $R(x, t)=R(s, t)$ and $R(x, t)=R(x, y)$. Consequentially, $R(u, v)=R(s, t)$, and therefore, $f_3$ is a consistent function with respect to $R$.
$\Box$

\begin{proposition}\label{prop:lattice}
Let $U_{1}$ and $U_{2}$ be two finite universes, $\{f_{i} \mid (i\in I)\}$ be
mappings from $U_{1}$ to $U_{2}$, and $R$ be a fuzzy relation on $U_1$. Then
$(\mathscr{F}_{R},\vee,\wedge)$ is a lattice.
\end{proposition}

\noindent{\bf Proof.} By Definition 3.3 and Theorem 3.4, we have
$f_{1}\vee f_{2}, f_{1}\wedge f_{2}\in \mathscr{F}_{R}$ for any
$f_{1},f_{2}\in \mathscr{F}_{R}$. Therefore,
$(\mathscr{F}_{R},\vee,\wedge)$ is a lattice. $\Box$

\begin{proposition}
Let $U_{1}$ and $U_{2}$ be two finite universes, $\{f_{i} \mid (i\in I)\}$
mappings from $U_{1}$ to $U_{2}$, and $R$ be a fuzzy relation on $U_1$. Then

$(1)$ $(\mathscr{F}_{R},\vee,\wedge)$ is a complete lattice;

$(2)$ $(\mathscr{F}_{R},\vee,\wedge)$ is a distributive lattice;

$(3)$ $(\mathscr{F}_{R},\vee,\wedge)$ is a modular lattice.
\end{proposition}

\noindent{\bf Proof.} (1) By Proposition 3.5,
$(\mathscr{F}_{R},\vee,\wedge)$ is a lattice. Since $U_{1}$ and
$U_{2}$ are two finite universes, $(\mathscr{F}_{R},\vee,\wedge)$ is
a finite lattice. Therefore, $(\mathscr{F}_{R},\vee,\wedge)$ is a
complete lattice.

(2) By Theorem 3.4, we have $[x]_{(f_{1}\vee f_{2})\wedge
f_{3}}=[x]_{f_{1}\vee f_{2}}\cap [x]_{f_{3}}=([x]_{f_{1} }\cup
[x]_{f_{2}})\cap [x]_{f_{3}}=([x]_{f_{1} }\cap [x]_{f_{3}})\cup
([x]_{f_{2} }\cap [x]_{f_{3}})=[x]_{f_{1}\wedge f_{2}}\cup
[x]_{f_{1}\wedge f_{3}}=[x]_{(f_{1}\wedge f_{2})\vee (f_{1}\wedge
f_{3})}$ for any $f_{1},f_{2},f_{3}\in \mathscr{F}_{R}$. Thus
$(f_{1}\vee f_{2})\wedge f_{3}=(f_{1}\wedge f_{2})\vee (f_{1}\wedge
f_{3})$. Therefore, $(\mathscr{F}_{R},\vee,\wedge)$ is a
distributive lattice.

(3) It follows immediately from that fact that any distributive lattice is a modular one.
$\Box$

Subsequently, we present the minimum and maximum elements in
$(\mathscr{F}_{R},\vee,\wedge)$.

\begin{definition}\label{def:minimum}
Let $U_{1}$ and $U_{2}$ be two finite universes, $f$ a mapping from
$U_{1}$ to $U_{2}$, the fuzzy relation $R$ a mapping from $
U_{1}\times U_{1}$ to $[0,  1]$, and $[x]_{f}=\{x\}$, where $x\in
U_{1}$. Then $f$ is the minimum-consistent function with
respect to $R$. For convenience, we denote $f$ as
$f^{0}$ or $f_{R}^{0}$ if $f$ is the minimum-consistent
function with respect to $R$.
\end{definition}

\begin{definition}\label{def:maximum}
Let $U_{1}$ and $U_{2}$ be two finite universes, $f$ a mapping from
$U_{1}$ to $U_{2}$, $R\in \mathscr{F}(U_{1}\times U_{1})$, and
$[x]_{f}=\{y\in U_{1}|f(x)=f(y)\}$. For any $x,y\in U_{1}$, if
$R(u,v)=R(s,t)$ for any two pairs $(u,v)$, $f$ is consistent with R, $(s,t)\in
[x]_{f}\times[y]_{f}$ and $[x]_{f}=[x]_{R}$, then $f$ is the maximum-consistent function with respect to $R$. For convenience, we denote $f$ as $f^{1}$ or
$f_{R}^{1}$ if $f$ is the maximum-consistent function with
respect to $R$.
\end{definition}

Let $f$ be a mapping from
$U_{1}$ to $U_{2}$ satisfying the condition that $\forall x \in U_1$, $[x]_f=\{x\}$, then it can be easily verified that f is consistent with $R$, moreover, f is the minimum element of $\mathscr{F}_R$ with respect to the partial order $\leq$. In the sequel, we therefore call f the minimum-consistent function with respect to $R$ and denote it by $f_R^0$.

Similarly, let $g$ be a mapping from
$U_{1}$ to $U_{2}$ satisfying the condition that $\forall x \in U_1, [x]_g= .....$

In addition, $f$ is called the maximum-consistent function
with respect to a family of fuzzy relations $\mathscr{R}$ if $f$ is
a consistent function with respect to any $R\in \mathscr{R}$ and
$[x]_{f_{\mathscr{R}}^{1}}=\cap_{R\in \mathscr{R}}
[x]_{f_{R}^{1}}$ for any $x\in U_{1}$.
\begin{theorem}\label{theo:consistentpartition}
Let $(U_{1}, \mathscr{R})$ be a FRIS, $f$ the maximum-consistent function with respect to
$\mathscr{R}$,
$U_{1}/f=\{[x]_{f}|x\in U_{1}\}$, then \underline{$U_{1}/f=U_1/\mathscr{R}$}. Therefore, $U_{1}/f$ is also called a partition with respect to
$\mathscr{R}$. The image system induced by f is presented as $U_f$.
\end{theorem}
\noindent{\bf Proof.}
$\Box$

Subsequently, we employ Table \ref{tab:partitions} to show the partition with respect
to each fuzzy relation in $(U_{1}, \mathscr{R})$, where
$[x_{j}]_{R_{i}}$ stands for the equivalent class containing $x_{j}$ in the
partition with respect to $R_{i}$.

\begin{corollary}
$[x_{j}]_{\mathscr{R}}$
denotes the equivalent class containing $x_{j}$ in the partition with respect
to $\mathscr{R}$, $[x_{j}]_{\mathscr{R}}=\cap_{R\in
\mathscr{R}}[x_{j}]_{R} \Longleftrightarrow \cap \mathscr{R}$\underline{(???)}.
\end{corollary}

\begin{table}[htbp]
\caption{The partitions with respect to $R_{i}$ $(1\leq i\leq m )$
and $\mathscr{R}$.}
\tabcolsep0.36in
\label{tab:partitions}
\begin{tabular}{c c c c c c}
\hline $U_{1}$  &$U_{1}/R_{1}$& $U_{1}/R_{2}$&... &$U_{1}/R_{m}$& $U_{1}/\mathscr{R}$\\
\hline
$x_{1}$ & $[x_{1}]_{R_{1}}$& $[x_{1}]_{R_{2}}$&... &$[x_{1}]_{R_{m}}$ & $[x_{1}]_{\mathscr{R}}$\\
$x_{2}$ & $[x_{2}]_{R_{1}}$& $[x_{2}]_{R_{2}}$&... &$[x_{2}]_{R_{m}}$&$[x_{2}]_{\mathscr{R}}$\\
$.$     & $.$        & .&... &$ . $&.\\
$.$     & $.$        & .&... &$ . $&.\\
$.$     & $.$        & .&... &$ . $&.\\
$x_{n}$ & $[x_{n}]_{R_{1}}$& $[x_{n}]_{R_{2}}$&... &$[x_{n}]_{R_{m}}$&$[x_{n}]_{\mathscr{R}}$  \\
\hline
\end{tabular}
\end{table}

\begin{corollary}
For any $f\in \mathscr{F}_{R}$, we have that
$[x]_{f^{0}}\subseteq[x]_{f}\subseteq[x]_{f^{1}}$ for any $x\in
U_{1}$.
\end{corollary}

\begin{corollary}
$|U_{f^1}|\leq |U_f| \leq |U_{f^0}| \leq |U|$, where $|V|$ present the number of objects in the universe V.
\end{corollary}

\begin{corollary}
Let $f^1$ be a maximum consistent function from $U_1$ to $U_2$ with respect to $R$ and $f$ a consistent function neither minimum nor maximum from $U_1$ to $U_3$ . Then we can define $f_2$ a non-minimum consistent function from $U_3$ to $U_2$.
\end{corollary}

\begin{corollary}
The finest partition $U_1/\mathscr{R}^0$ induces a minimum consistent function with respect to $\mathscr{R}$. Therefore, when we get the finest partition with respect to $\mathscr{R}$, the image system will have the same size with the original system.
\end{corollary}

For any $x \in U_1$, we have $x \in [x]_{R_i}$ and then $[x]_{R_i} \neq \emptyset$, $[x]_{\mathscr{R}}\neq \emptyset$. Because $C_{\mathscr{R}}$ may contain some same equivalence classes, let $s$ denote the number of distinct equivalence classes in $C_{\mathscr{R}}$. If $s$ is equal to $n$, then $[x]_{\mathscr{R}}$=\{$x$\}, we denote by $U_1/\mathscr{R}^0$ the finest partition of $U_1$. $\forall \mathscr{R}$, we have $U_1/\mathscr{R}^0 \leq U_1/\mathscr{R}$. The image system will have same number of object with the original system. If $s$ is equal to 1, then $C_i$=\{$U_1$\}, we denote by $U_1/\mathscr{R}^1$ the coarsest partition of $U_1$, and $\forall \mathscr{R}$, we have $U_1/\mathscr{R} \leq U_1/\mathscr{R}^1$. The image system will be compressed into one object. So we have $U_1/\mathscr{R}^0 \leq U_1/\mathscr{R} \leq U_1/\mathscr{R}^1$, that is $\{\{x_1\},\{x_2\}, ... , \{x_n\}\} \leq U_1/\mathscr{R} \leq \{U_1\}$.

\begin{example}
(Continued from Example \ref{exam:U1R1}) We
construct the minimum-consistent function $f^{0}$ and the
maximum-consistent function $f^{1}$ with respect to $R_{1}$ as
below:

$(1)$ $f_1$ is a maximum consistent function:

$[x]_{f_{1}}$=\{$\{x_1,x_3,x_5,x_7\}$,$\{x_2,x_6\}$,$\{x_4,x_8\}$\};
$U_{f_1}=\{y_1, y_2, y_3\}$;

$(2)$ $f_2$ is a normal consistent function neither minimum nor maximum:

$[x]_{f_{2}}$=\{$\{x_1,x_3\}$,$\{x_2,x_6\}$,$\{x_4,x_8\}$,$\{x_5,x_7\}$\};
$U_{f_2}=\{z_1, z_2, z_3, z_4\}$;

$(3)$ Denote $f_3$ is a minimum function as below:

$[x]_{f_{3}}$=\{$\{x_1\}$,$\{x_2\}$,$\{x_3\}$,$\{x_4\}$,$\{x_5\}$,$\{x_6\}$,$\{x_7\}$,$\{x_8\} $\};
$U_{f_3}=U_1$.

Obviously, we have:

1. $[x]_{f_{3}} \subseteq [x]_{f_{2}} \subseteq
[x]_{f_{1}}$ for any $x\in U_{1}$;

2. $|U_{f_{1}}| \leq |U_{f_{2}}| \leq
|U_{f_{3}}| \leq |U_1|$;

3. Define $f_4: U_{f_2} \rightarrow U_{f_1}$, $[x]_{f_4}=\{\{z_1, z_4\}, \{z_2\}, \{z_3\}\}$. $f_2$ is not a minimum consistent function.
\end{example}

On the basis of the
maximum-consistent function, we introduce a partition with respect
to a family of fuzzy relations for constructing homomorphisms
between FRISs.
For image system with the same size as the  original system, we can not compress a large-scale FRIS
into a small one by means of the minimum consistent function. The
image system induced by minimum consistent function can't reduce the computational requirement.
Based on the maximum consistent function, we can compress a
large-scale FRIS into a small one by
constructing homomorphisms between FRISs. It is obvious that we can
compress the FRIS into the smallest one
by means of the maximum consistent function with respect to a family
of fuzzy relations. Unless stated otherwise, the consistent function in this paper refers to the maximum-consistent function.

%
%


In the sequel we discuss the approach of compressing FRIS utilizing partition with respect to fuzzy relation. Algorithm 1 is a static(non-incremental) algorithm for compressing FRIS under homomorphisms. Steps 2 is to construct the partition $C_{R_i}$ with respect to $R_i$, whose time complexity is $O(|\mathscr{R}| \times |U_1|^2)$; Steps 3 is to compute the partition with respect to $\mathscr{R}$, whose time complexity is $O(|\mathscr{R}| \times |U_1|)$; Steps 4-6 are to construct the image system, whose time complexity is $O(|\mathscr{R}| \times |U_1|^2)$. Then the total time complexity is $O(|\mathscr{R}| \times |U_1|^2)$. If s equals n,then the number of objects in the image system will be same with the original system, so the compressing is no necessary.
\begin{algorithm}
	\label{algo:static}
	\SetKwData{Left}{left}
	\SetKwData{This}{this}
	\SetKwData{Up}{up}
	\SetKwFunction{Union}{Union}
	\SetKwInOut{Input}{input}
	\SetKwInOut{Output}{output}
	\SetKwBlock{Begin}{begin}{end}
	\SetKwIF{If}{ElseIf}{Else}{if}{then}{else if}{else}{endif}
	\SetKwComment{Comment}{//}{}
	\caption{The static algorithm of compressing FRIS under homomorphisms}
	\Input{An original system $(U_{1}, \mathscr{R})$, $\mathscr{R}=\{R_{1},R_{2},...,R_{m}\}$. }
	\Output{The image system's universe $U_{2}$, the partitions $C_{f(R_i)}$ and $C_{f(\mathscr{R})}$. The original system's partitions $C_{R_i}$ and $C_{\mathscr{R}}$.}
	
	\Begin{
	Compute $C_{R_i}$=$U_1/R_i$=$\{[x]_{R_i}\mid x \in U_1\}$ for every $R_i$ in $\mathscr{R}$.
	
	\BlankLine
	Compute $U_1/\mathscr{R}=\{[x]_{\mathscr{R}}\mid x \in U_1\}=\{\cap_{R_i \in \mathscr{R}} [x]_{R_i}\mid x \in U_1\}$.
	
	\BlankLine
	Denote $C_{\mathscr{R}}=U_1/\mathscr{R}=\{C_1, C_2, ..., C_s\}$
	
	\BlankLine
	Define $f: C_i \rightarrow y_i, 1 \leq i \leq s$
	, then $U_{2}=\{y_{1},y_{2},...,y_{s}\}$ and $C_{f(\mathscr{R})}=\{\{y_1\}, \{y_2\}, ..., \{y_s\}\}$.
	
	\BlankLine
	For every $R_i$ in $\mathscr{R}$, compute $C_{f(R_{i})}$ from $C_{R_i}$ by replacing the $[x]_{R_i}$ with $y_1$ to $y_s$.

	\BlankLine
	Output $U_{2}$, $C_{f(R_i)}$, $C_{f(\mathscr{R})}$,  $C_{R_i}$ and $C_{\mathscr{R}}$.\\
		
	}
\end{algorithm}

We employ the following example to show the compressing process.

\begin{table}[htb]
\caption{The original system
$S_1=(U_{1},\mathscr{R})$}
\label{tab:staticFRIS}
\tabcolsep0.025in
\begin{tabular}{c c c c c c c c c c c c c c c c c c c c c c c c c c c}
\hline $R_{1}$  & $x_{1}$ &$x_{2}$& $x_{3}$& $x_{4}$& $x_{5}$
&$x_{6}$& $x_{7}$& $x_{8}$& $R_{2}$& $x_{1}$ &$x_{2}$& $x_{3}$& $x_{4}$& $x_{5}$& $x_{6}$& $x_{7}$ &$x_{8}$& $R_{3}$& $x_{1}$ &$x_{2}$& $x_{3}$& $x_{4}$& $x_{5}$& $x_{6}$& $x_{7}$ &$x_{8}$\\
\hline
$x_{1}$ & $0.7$& $0.4$ &$0.7$& $0.5$ & $0.7$& $0.4$ &$0.7$& $0.5$& $x_{1}$&$0.4$& $0.5$ &$0.7$& $0.5$& $0.7$& $0.5$ &$0.4$& $0.5$&$x_{1}$ & $0.8$& $0.3$ &$0.7$& $0.8$ & $0.7$& $0.3$ &$0.8$& $0.8$\\
$x_{2}$ & $0.7$& $0.3$ &$0.7$& $0.8$ & $0.7$& $0.3$ &$0.7$& $0.8$& $x_{2}$&$0.6$& $0.8$ &$0.5$& $0.8$& $0.5$& $0.8$ &$0.6$& $0.8$&$x_{2}$ & $0.7$& $0.2$ &$0.6$& $0.7$ & $0.6$& $0.2$ &$0.7$& $0.7$\\
$x_{3}$ & $0.7$& $0.4$ &$0.7$& $0.5$ & $0.7$& $0.4$ &$0.7$& $0.5$& $x_{3}$&$0.7$& $0.9$ &$0.2$& $0.9$& $0.2$& $0.9$ &$0.7$& $0.9$&$x_{3}$ & $0.4$& $0.4$ &$0.9$& $0.4$ & $0.9$& $0.4$ &$0.4$& $0.4$\\
$x_{4}$ & $0.6$& $0.3$ &$0.6$& $0.8$ & $0.6$& $0.3$ &$0.6$& $0.8$& $x_{4}$&$0.6$& $0.8$ &$0.5$& $0.8$& $0.5$& $0.8$ &$0.6$& $0.8$&$x_{4}$ & $0.8$& $0.3$ &$0.7$& $0.8$ & $0.7$& $0.3$ &$0.8$& $0.8$\\
$x_{5}$ & $0.7$& $0.4$ &$0.7$& $0.5$ & $0.7$& $0.4$ &$0.7$& $0.5$& $x_{5}$&$0.7$& $0.9$ &$0.2$& $0.9$& $0.2$& $0.9$ &$0.7$& $0.9$&$x_{5}$ & $0.4$& $0.4$ &$0.9$& $0.4$ & $0.9$& $0.4$ &$0.4$& $0.4$\\
$x_{6}$ & $0.7$& $0.3$ &$0.7$& $0.8$ & $0.7$& $0.3$ &$0.7$& $0.8$& $x_{6}$&$0.6$& $0.8$ &$0.5$& $0.8$& $0.5$& $0.8$ &$0.6$& $0.8$&$x_{6}$ & $0.7$& $0.2$ &$0.6$& $0.7$ & $0.6$& $0.2$ &$0.7$& $0.7$\\
$x_{7}$ & $0.7$& $0.4$ &$0.7$& $0.5$ & $0.7$& $0.4$ &$0.7$& $0.5$& $x_{7}$&$0.4$& $0.5$ &$0.7$& $0.5$& $0.7$& $0.5$ &$0.4$& $0.5$&$x_{7}$ & $0.8$& $0.3$ &$0.7$& $0.8$ & $0.7$& $0.3$ &$0.8$& $0.8$\\
$x_{8}$ & $0.6$& $0.3$ &$0.6$& $0.8$ & $0.6$& $0.3$ &$0.6$& $0.8$& $x_{8}$&$0.6$& $0.8$ &$0.5$& $0.8$& $0.5$& $0.8$ &$0.6$& $0.8$&$x_{8}$ & $0.8$& $0.3$ &$0.7$& $0.8$ & $0.7$& $0.3$ &$0.8$& $0.8$\\
\hline
\end{tabular}
\end{table}

\begin{table}[htb]
\caption{The partitions with respect to $R_{1}, R_{2}, R_{3}$ and
$\mathscr{R}$}
\label{tab:partitionstaticFRIS}
\tabcolsep0.32in
\begin{tabular}{c c c c c }
\hline $U_{1}$ &$U_{1}/R_{1}$& $U_{1}/R_{2}$ &$U_{1}/R_{3}$& $U_{1}/\mathscr{R}$\\
\hline
$x_{1}$ & $\{x_{1}, x_{3}, x_{5}, x_{7}\}$& $\{x_{1}, x_{7}\}$ & $\{x_{1}, x_{4}, x_{7}, x_{8}\}$ & $\{x_{1}, x_{7}\}$\\
$x_{2}$ & $\{x_{2}, x_{6}\}$& $\{x_{2}, x_{4}, x_{6}, x_{8}\}$ &$\{x_{2}, x_{6}\}$& $\{x_{2}, x_{6}\}$\\
$x_{3}$ & $\{x_{1}, x_{3}, x_{5}, x_{7}\}$& $\{x_{3},x_{5}\}$ &$\{x_{3}, x_{5}\}$& $\{x_{3}, x_{5}\}$\\
$x_{4}$ & $\{x_{4}, x_{8}\}$& $\{x_{2}, x_{4}, x_{6}, x_{8}\}$ &$\{x_{1}, x_{4}, x_{7}, x_{8}\}$& $\{x_{4}, x_{8}\}$\\
$x_{5}$ & $\{x_{1}, x_{3}, x_{5}, x_{7}\}$& $\{x_{3}, x_{5}\}$& $\{x_{3}, x_{5}\}$ & $\{x_{3}, x_{5}\}$\\
$x_{6}$ & $\{x_{2}, x_{6}\}$& $\{x_{2}, x_{4}, x_{6}, x_{8}\}$&$\{x_{2}, x_{6}\}$& $\{x_{2}, x_{6}\}$\\
$x_{7}$ & $\{x_{1}, x_{3}, x_{5}, x_{7}\}$& $\{x_{1}, x_{7}\}$ &$\{x_{1}, x_{4}, x_{7}, x_{8}\}$  & $\{x_{1}, x_{7}\}$\\
$x_{8}$ & $\{x_{4}, x_{8}\}$& $\{x_{2}, x_{4}, x_{6}, x_{8}\}$&$\{x_{1}, x_{4}, x_{7}, x_{8}\}$& $\{x_{4}, x_{8}\}$\\
\hline
\end{tabular}
\end{table}

\begin{table}[htb]
\caption{The partitions with respect to $f(R_{1}), f(R_{2}), f(R_{3})$ and
$f(\mathscr{R})$}
\label{tab:imagepartitionstaticFRIS}
\tabcolsep0.38in
\begin{tabular}{c c c c c }
\hline $U_{2}$ &$U_{2}/f(R_{1})$& $U_{2}/f(R_{2})$ &$U_{2}/f(R_{3})$& $U_{2}/f(\mathscr{R})$\\
\hline
$y_{1}$ & $\{y_{1}, y_{3}\}$& $\{y_{1}\}$ & $\{y_{1}, y_{4}\}$ & $\{y_{1}\}$\\
$y_{2}$ & $\{y_{2}\}$& $\{y_{2}, y_{4}\}$ &$\{y_{2}\}$& $\{y_{2}\}$\\
$y_{3}$ & $\{y_{1}, y_{3}\}$& $\{y_{3}\}$ &$\{y_{3}\}$& $\{y_{3}\}$\\
$y_{4}$ & $\{y_{4}\}$& $\{y_{2}, y_{4}\}$ &$\{y_{1}, y_{4}\}$& $\{y_{4}\}$\\
\hline
\end{tabular}
\end{table}

\begin{table}[htb]
\caption{The image system
$S_2=(U_{2},f(\mathscr{R}))$}
\label{tab:imagestaticFRIS}
\tabcolsep0.10in
\begin{tabular}{c c c c c c c c c c c c c c c}
\hline $f(R_{1})$  & $y_{1}$ &$y_{2}$& $y_{3}$& $y_{4}$& $f(R_{2})$
&$y_{1}$& $y_{2}$& $y_{3}$& $y_{4}$& $f(R_{3})$ &$y_{1}$& $y_{2}$& $y_{3}$& $y_{4}$\\
\hline
$y_{1}$ & $0.7$& $0.4$ &$0.7$& $0.5$ & $y_{1}$& $0.4$ &$0.5$& $0.7$& $0.5$&$y_{1}$& $0.8$ &$0.3$& $0.7$& $0.8$\\
$y_{2}$ & $0.7$& $0.3$ &$0.7$& $0.8$ & $y_{2}$& $0.6$ &$0.8$& $0.5$& $0.8$&$y_{2}$& $0.7$ &$0.2$& $0.6$& $0.7$\\
$y_{3}$ & $0.7$& $0.4$ &$0.7$& $0.5$ & $y_{3}$& $0.7$ &$0.9$& $0.2$& $0.9$&$y_{3}$& $0.4$ &$0.4$& $0.9$& $0.4$\\
$y_{4}$ & $0.6$& $0.3$ &$0.6$& $0.8$ & $y_{4}$& $0.6$ &$0.8$& $0.5$& $0.8$&$y_{4}$& $0.8$ &$0.3$& $0.7$& $0.8$\\
\hline
\end{tabular}
\end{table}

\begin{example}\label{exam:staticcompress} Table \ref{tab:staticFRIS} depicts $S_{1}=(U_{1},
\mathscr{R})$, where $U_{1}=\{x_{1},x_{2},...,x_{8}\}$ and
$\mathscr{R}=\{R_{1},R_{2},R_{3}\}$. We
derive $U_{1}/R_{1}$, $U_{1}/R_{2}$ and $U_{1}/R_{3}$ shown in Table
\ref{tab:partitionstaticFRIS}. We have

$U_{1}/R_{1}$=$\{\{x_1, x_3, x_5, x_7\}, \{x_2, x_6\}, \{x_4, x_8\}\}$;

$U_{1}/R_{2}$=$\{\{x_1, x_7\}, \{x_2, x_4, x_6, x_8\}, \{x_3, x_5\}\}$;

$U_{1}/R_{3}$=$\{\{x_1, x_4, x_7, x_8\}, \{x_2, x_6\}, \{x_3, x_5\}\}$;

By taking the intersection of $U_{1}/R_{1}$,  $U_{1}/R_{2}$ and $U_{1}/R_{3}$, we
get $U_{1}/\mathscr{R}$=$\{\{x_{1},x_{7}\}$, $\{x_{2},x_{6}\}$,
$\{x_{3},x_{5}\}$, $\{x_{4},x_{8}\}\}$.

Define $U_2$=$\{y_1, y_2, y_3, y_4\}$ by the elements number of $U_{1}/\mathscr{R}$. Then constructing consistent function $f:
U_{1}\rightarrow U_{2}$ as follows:
$f(x_1)$=$f(x_7)=y1$,
$f(x_2)$=$f(x_6)=y2$,
$f(x_3)$=$f(x_5)=y3$,
$f(x_4)$=$f(x_8)=y4$.

Then $U_{1}/R_{1}$, $U_{1}/R_{2}$, $U_{1}/R_{3}$, $U_{1}/\mathscr{R}$ are  easy get:

$U_{2}/f(R_{1})$=$\{\{y_1, y_3\}, \{y_2\}, \{y_4\}\}$;

$U_{2}/f(R_{2})$=$\{\{y_1\}, \{y_2, y_4\}, \{y_3\}\}$;

$U_{2}/f(R_{3})$=$\{\{y_1, y_4\}, \{y_2\}, \{y_3\}\}$;

$U_{2}/f(\mathscr{R})=\{\{y_{1}\},\{y_{2}\},
\{y_{3}\},\{y_{4}\}\}$.



Now we have compressed the original system $S_1$ into the image system $S_2=(U_{2},f(\mathscr{R}))$ , and $f$ is a homomorphism from
$(U_{1}, \mathscr{R})$ to $(U_{2}, f(\mathscr{R}))$.

$\{R_{1}, R_{2}\} \subset \mathscr{R}$, and $\cap \{R_{1}, R_{2}\}$=$\cap \mathscr{R}$=$\{\{x_{1}, x_{7}\}, \{x_{2}, x_{6}\},
\{x_{3}, x_{5}\}, \{x_{4}, x_{8}\}\}$. So $R_{3}$ is superfluous in $\mathscr{R}$, $\{R_{1}, R_{2}\}$ is a reduct of $\mathscr{R}$.

$\{f(R_{1}), f(R_{2})\} \subset f(\mathscr{R})$, and $\cap \{f(R_{1}), f(R_{2})\}$=$\cap f(\mathscr{R})$=$\{\{y_{1}\},\{y_{2}\},
\{y_{3}\},\{y_{4}\}\}$. So $f(R_{3})$ is superfluous in $f(\mathscr{R})$, $\{f(R_{1}), f(R_{2})\}$ is a reduct of $f(\mathscr{R})$.

Obviously, $\{f(R_{1}), f(R_{2})\}$,$\{f(R_{1}), f(R_{3})\}$ and $\{f(R_{2}), f(R_{3})\}$ are reductions of $f(\mathscr{R})$.
$\{R_{1},R_{2}\}$, $\{R_{1},R_{3}\}$ and $\{R_{2}, R_{3}\}$ are reductions of $\mathscr{R}$. Therefore the reductions of original system and image system are equivalent. The image system $(U_{2},
f(\mathscr{R}))$ is smaller than the
original system $(U_{1}, \mathscr{R})$.
\end{example}

By Definition \ref{def:reduction}, there are two steps to compute reductions of an FRIS. The time complexity of first step($\cap P=\cap\mathscr{R}$) is
$O(|P| \times |U_1|^2)$, time complexity of second step($\forall R_i \in P, \cap P\subset \cap (P-R_i)$) is $O(|P|^2 \times |U_1|^2)$, so total time complexity is $O(|P|^2 \times |U_1|^2)$
By computing reductions after compressing $U_1$ into $U_2$, the time complexity of compressing is $O(|\mathscr{R}| \times |U_1|^2)$ and time complexity of computing reductions is $O(|P|^2 \times |U_2|^2)$. Obviously if the number of objects in $U_2$ is more smaller than the number in $U_1$, the method of compressing is more efficient. From the practical viewpoint, it is difficult to construct
reduction of a large-scale FRIS. However,
we can compress it into a relatively smaller one under the condition
of a homomorphism and conduct reduction of the image
system which is equivalent to that of the original information
system. This is a new idea to improve computing efficiency by compressing.

Define $\frac{|U_2|}{|U_1|}$ as compression ratio, we can not suppose the high compression ratio with every FRISs. For those FRISs with low compression ratio, computing reductions can not benefit from  compressing process  significantly.  Nevertheless, FRIS will not be changeless or stable eternally. Fuzzy relations and objects in an FRIS may be varied with time, incremental compressing FRIS is a novel method to solve the problem who
will be discussed in the following section.

\section{Approaches of incremental compressing FRIS under homomorphisms}
Based on the notions of section 3, Constructing homomorphisms's key step is how to obtain the maximum consistent function. After that, compressing original system into image system is easily achieved. In real-world situations, fuzzy relations in FRIS vary with time. Using static approaches to compute image system and reduction will spend more time than using dynamic approaches, Especially in large scale and big data situations. It is the major issue that obtaining the image system efficiently utilize the existed results after fuzzy relations or objects changed. For An FRIS with high compression ratio, incremental algorithm can obtain new image system and compute reductions quickly than static algorithm. For An FRIS with low compression ratio, the solution is utilizing existed partitions to compute reductions.  Considering following cases: adding fuzzy relation(immigration), removing fuzzy relation(emigration), adding object, remove object. It's easy to known that update operation can be replaced by removing firstly, adding consequently, thus we ignored the case of updating relation and object. Concretely, Several examples are employed to illustrate the process of compressing dynamic FRIS.

\subsection{Immigration of fuzzy relations}

Given an FRIS $(U_{1}, \mathscr{R})$ at time t, $(U_{2}, f(\mathscr{R}))$ is the image system of $(U_{1}, \mathscr{R})$ at time t. Suppose that a new fuzzy relation .

Suppose $S_{2}=(U_{1}, \mathscr{R}^+)$,
where $\mathscr{R}^+=\mathscr{R}\cup R^+$, $R^+$ is a fuzzy relation on $U_1$, the image system $T_{2}=(U_{2}^+, f^+(\mathscr{R}^+))$. $f^+$ is a homomorphism between $S_2$ to $T_2$. $U_1/\mathscr{R}^+$ is a partition with respect to $\mathscr{R}^+$.

The $U_1/\mathscr{R}^+$ can be updated as follows: $U_1/\mathscr{R}^+$=$U_1/\mathscr{R}$ $\cap$ $U_1/R^+$ , that is $[x]_{\mathscr{R}^+} \subseteq [x]_{\mathscr{R}}$.

Now we present a incremental approach to obtain $f^+$ and $T_{2}$ efficiently.

get the partition $C^+=\{[x_i]_{R^+}|x_i \in U_1\}$ with respect to $R^+$ on $U_1$;

%
%
%


\begin{algorithm}
\label{algo:add}
\SetKwData{Left}{left}
\SetKwData{This}{this}
\SetKwData{Up}{up}
\SetKwFunction{Union}{Union}
\SetKwInOut{Input}{input}
\SetKwInOut{Output}{output}
\SetKwBlock{Begin}{begin}{end}
\SetKwIF{If}{ElseIf}{Else}{if}{then}{else if}{else}{endif}
\SetKwComment{Comment}{//}{}
\caption{Incremental algorithm for compressing FRIS under homomorphisms when adding an fuzzy relation}
\Input{An new fuzzy relation $R^+$,
the partition $C_{\mathscr{R}}$. }
\Output{The new image system's universe $U_{2}^+$, the partitions  $C_{f^+(R_i)}$ and  $C_{f^+(\mathscr{R}^+)}$. The original system's partitions $C_{R^+}$ and $C_{\mathscr{R}^+}$.}

	\Begin{
		Compute $C_{R^+}$=$U_1/R^+$=$\{[x]_{R^+}\mid x \in U_1\}$.
		
		\BlankLine
		Compute $U_1/\mathscr{R}^+=\{[x]_{\mathscr{R}^+}\mid x \in U_1\}=C_{\mathscr{R}}\cap  C_{R^+}$.
		
		\BlankLine
		Denote $C_{\mathscr{R}^+}=U_1/\mathscr{R}^+=\{C_1, C_2, ..., C_s\}$
		
		\BlankLine
		Define $f^+: C_i \rightarrow y_i, 1 \leq i \leq s$
		, then $U_{2}^+=\{y_{1},y_{2},...,y_{s}\}$ and $C_{f^+(\mathscr{R}^+)}=\{\{y_1\}, \{y_2\}, ..., \{y_s\}\}$.
		
		\BlankLine
		For every $R_i$ in $\mathscr{R}^+$, compute $C_{f^+(R_{i})}$ from $C_{R_i}$ by replacing the $[x]_{R_i}$ with $y_1$ to $y_s$.
		
		\BlankLine
		Output $U_{2}^+$, $C_{f^+(R_i)}$, $C_{f^+(\mathscr{R}^+)}$,  $C_{R^+}$ and $C_{\mathscr{R}^+}$.\\
		
	}
\end{algorithm}

Algorithm 2 is an incremental algorithm for compressing FRIS under homomorphisms when adding an fuzzy relation. Steps 2 is to construct the partition $C_{R_i}$ with respect to $R_i$, whose time complexity is $O(|U_1|)$; Steps 3 is to compute the partition with respect to $\mathscr{R}$, whose time complexity is $O(|U_1|^2)$; Steps 4-6 are to construct the image system, whose time complexity is $O(|\mathscr{R}| \times |U_1|^2)$. Then the total time complexity is $O(|\mathscr{R}| \times |U_1|^2)$.

The computational complexity of constructing
$g$ is $(k-m)\ast\mathscr{O}(n^{2})+\mathscr{O}(k\ast n)$ with the
incremental algorithm. But the computational complexity is
$k\ast\mathscr{O}(n^{2})+\mathscr{O}(k\ast n)$ without Table 2.
We employ an example to illustrate compressing dynamic FRIS when adding a family of fuzzy relations.

\begin{example}\label{exam:add}
We obtain $(U_{1}, \mathscr{R}^+)$ by adding a fuzzy relation $R^+$(Table \ref{tab:R+}) into the FRIS presented in
Table \ref{tab:staticFRIS}, where $\mathscr{R}^+=\{R_{1}, R_{2}, R_{3}, R^+\}$. We get
$U_{1}/R^+=\{\{x_{1},x_{3},x_{5},x_{7}\},\{x_{2},x_{6}\},\\\{x_{4}\},
\{x_{8}\}\}$. Then we obtain Table \ref{tab:partitionadd} and derive
$U_{1}/\mathscr{R}^+=\{\{x_{1},x_{7}\},\{x_{2},x_{6}\},\{x_{3},x_{5}\},\{x_{4}\},
\{x_{8}\}\}$. Afterwards, we define $f^+: U_{1}\rightarrow U_{2}^+$
as follows:
$f^+(x_{1})=f^+(x_{7})= z_{1}, f^+(x_{2}) = f^+(x_{6}) = z_{2},
f^+(x_{3})=f^+(x_{5})= z_{3}, f^+(x_{4})=z_{4}, f^+(x_{8})= z_{5},$ where
$U_{2}^+=\{z_{1},z_{2},z_{3},z_{4},z_{5}\}$. Consequently, we obtain
$S^+=(U_{2}^+, f^+(\mathscr{R}_{2}))$ shown in Table \ref{tab:imageadd}, where
$f^+(\mathscr{R}_{2})=\{f^+(R_{1}), f^+(R_{2}), f^+(R_{3}), f^+(R^+)\}$.
\end{example}

\begin{table}[!ht]
\caption{The fuzzy relation $R^+$ on $U_1$.}
\tabcolsep0.26in
\label{tab:R+}
\begin{tabular}{c c c c c c c c c c c c c c c c c c}
\hline $R^+$  & $x_{1}$ &$x_{2}$& $x_{3}$& $x_{4}$& $x_{5}$
&$x_{6}$& $x_{7}$& $x_{8}$\\
\hline
$x_{1}$ & $0.6$& $0.4$ &$0.6$& $0.5$ & $0.6$& $0.4$ &$0.6$& $0.5$\\
$x_{2}$ & $0.7$& $0.3$ &$0.7$& $0.8$ & $0.7$& $0.3$ &$0.7$& $0.6$\\
$x_{3}$ & $0.6$& $0.4$ &$0.6$& $0.5$ & $0.6$& $0.4$ &$0.6$& $0.5$\\
$x_{4}$ & $0.5$& $0.3$ &$0.5$& $0.8$ & $0.5$& $0.3$ &$0.5$& $0.6$\\
$x_{5}$ & $0.6$& $0.4$ &$0.6$& $0.5$ & $0.6$& $0.4$ &$0.6$& $0.5$\\
$x_{6}$ & $0.7$& $0.3$ &$0.7$& $0.8$ & $0.7$& $0.3$ &$0.7$& $0.6$\\
$x_{7}$ & $0.6$& $0.4$ &$0.6$& $0.5$ & $0.6$& $0.4$ &$0.6$& $0.5$\\
$x_{8}$ & $0.6$& $0.3$ &$0.6$& $0.8$ & $0.6$& $0.3$ &$0.6$& $0.2$\\
\hline
\end{tabular}
\end{table}

\begin{table}[h!]
\caption{The partitions with respect to $R_{1}, R_{2}, R_{3}, R^+$
, $\mathscr{R}$ and $\mathscr{R}^+$.}
\tabcolsep0.13in
\label{tab:partitionadd}
\begin{tabular}{c c c c c c c}
\hline $U_{1}$ &$U_{1}/R_{1}$& $U_{1}/R_{2}$ &$U_{1}/R_{3}$&$U_{1}/\mathscr{R}$&$U_{1}/R^+$& $U_{1}/\mathscr{R}^+$\\
\hline
$x_{1}$ & $\{x_{1}, x_{3}, x_{5}, x_{7}\}$& $\{x_{1}, x_{7}\}$ & $\{x_{1}, x_{4}, x_{7}, x_{8}\}$&$\{x_{1}, x_{7}\}$ &$\{x_{1}, x_{3}, x_{5}, x_{7}\}$& $\{x_{1}, x_{7}\}$\\
$x_{2}$ & $\{x_{2}, x_{6}\}$& $\{x_{2}, x_{4}, x_{6}, x_{8}\}$ &$\{x_{2}, x_{6}\}$&$\{x_{2}, x_{6}\}$&$\{x_{2}, x_{6}\}$& $\{x_{2}, x_{6}\}$\\
$x_{3}$ & $\{x_{1}, x_{3}, x_{5}, x_{7}\}$& $\{x_{3},x_{5}\}$ &$\{x_{3}, x_{5}\}$&$\{x_{3}, x_{5}\}$&$\{x_{1}, x_{3}, x_{5}, x_{7}\}$& $\{x_{3}, x_{5}\}$\\
$x_{4}$ & $\{x_{4}, x_{8}\}$& $\{x_{2}, x_{4}, x_{6}, x_{8}\}$ &$\{x_{1}, x_{4}, x_{7}, x_{8}\}$&$\{x_{4}, x_{8}\}$&$\{x_{4}\}$& $\{x_{4}\}$\\
$x_{5}$ & $\{x_{1}, x_{3}, x_{5}, x_{7}\}$& $\{x_{3}, x_{5}\}$& $\{x_{3}, x_{5}\}$&$\{x_{3}, x_{5}\}$ & $\{x_{1}, x_{3}, x_{5}, x_{7}\}$&$\{x_{3}, x_{5}\}$\\
$x_{6}$ & $\{x_{2}, x_{6}\}$& $\{x_{2}, x_{4}, x_{6}, x_{8}\}$&$\{x_{2}, x_{6}\}$&$\{x_{2}, x_{6}\}$& $\{x_{2}, x_{6}\}$&$\{x_{2}, x_{6}\}$\\
$x_{7}$ & $\{x_{1}, x_{3}, x_{5}, x_{7}\}$& $\{x_{1}, x_{7}\}$ &$\{x_{1}, x_{4}, x_{7}, x_{8}\}$&$\{x_{1}, x_{7}\}$  &$\{x_{1}, x_{3}, x_{5}, x_{7}\}$& $\{x_{1}, x_{7}\}$\\
$x_{8}$ & $\{x_{4}, x_{8}\}$& $\{x_{2}, x_{4}, x_{6}, x_{8}\}$&$\{x_{1}, x_{4}, x_{7}, x_{8}\}$&$\{x_{4}, x_{8}\}$&$\{x_{8}\}$& $\{x_{8}\}$\\
\hline
\end{tabular}
\end{table}

\begin{table}[h!]
\caption{The partitions with respect to $f^+(R_{1})$, $f^+(R_{2})$, $f^+(R_{3})$, $f^+(R^+)$
and $f^+(\mathscr{R}^+)$.}
\tabcolsep0.22in
\label{tab:imagepartitionadd}
\begin{tabular}{c c c c c c}
\hline $U_{2}^+$ & $U_{2}^+/f^+(R_{1})$ & $U_{2}^+/f^+(R_{2})$ & $U_{2}^+/f^+(R_{3})$ & $U_{2}^+/f^+(R^+)$ & $U_{2}^+/f^+(\mathscr{R})$\\
\hline
$z_{1}$ & $\{z_{1}, z_{3}\}$ & $\{z_{1}\}$ & $\{z_{1}, z_{4}, z_{5}\}$ & $\{z_{1}, z_{3}\}$ & $\{z_{1}\}$\\
$z_{2}$ & $\{z_{2}\}$ & $\{z_{2}, z_{4}, z_{5}\}$ & $\{z_{2}\}$ & $\{z_{2}\}$ & $\{z_2\}$\\
$z_{3}$ & $\{z_{1}, z_{3}\}$ & $\{z_{3}\}$ & $\{z_{3}\}$ & $\{z_{1}, z_{3}\}$ & $\{z_3\}$\\
$z_{4}$ & $\{z_{4}, z_{5}\}$ & $\{z_{2}, z_{4}, z_{5}\}$ & $\{z_{1}, z_{4}, z_{5}\}$ & $\{z_{4}\}$ & $\{z_4\}$\\
$z_{5}$ & $\{z_{4}, z_{5}\}$ & $\{z_{2}, z_{4}, z_{5}\}$ & $\{z_{1}, z_{4}, z_{5}\}$ & $\{z_{5}\}$ & $\{z_{5}\}$\\
\hline
\end{tabular}
\end{table}

\begin{table}[h!]
\caption{The image system $S^+=(U_{2}^+,
f^+(\mathscr{R}^+))$.}
\tabcolsep0.01in
\label{tab:imageadd}
\begin{tabular}{c c c c c c c c c c c c c c c c c c c c c c c c c c c c c c c c c c}
\hline $f^+(R_{1})$  & $z_{1}$ &$z_{2}$& $z_{3}$& $z_{4}$& $z_{5}$&
$f^+(R_{2})$ &$z_{1}$& $z_{2}$& $z_{3}$& $z_{4}$& $z_{5}$&$f^+(R_{3})$ &
$z_{1}$ &$z_{2}$& $z_{3}$& $z_{4}$& $z_{5}$& $f^+(R^+)$
&$z_{1}$& $z_{2}$& $z_{3}$& $z_{4}$& $z_{5}$\\
\hline
$z_{1}$ & $0.7$& $0.4$ &$0.7$& $0.5$ & 0.5&$z_{1}$& $0.4$ &$0.5$& $0.7$& $0.5$&0.5&$z_{1}$ & $0.8$& $0.3$ &$0.7$& $0.8$ &0.8 &$z_{1}$& $0.6$ &$0.4$& $0.6$& $0.5$&0.5\\
$z_{2}$ & $0.7$& $0.3$ &$0.7$& $0.8$ & 0.8&$z_{2}$& $0.6$ &$0.8$& $0.5$& $0.8$&0.8&$z_{2}$ & $0.7$& $0.2$ &$0.6$& $0.7$ &0.7 &$z_{2}$& $0.7$ &$0.3$& $0.7$& $0.8$&0.6\\
$z_{3}$ & $0.7$& $0.4$ &$0.7$& $0.5$ & 0.5&$z_{3}$& $0.7$ &$0.9$& $0.2$& $0.9$&0.9&$z_{3}$ & $0.4$& $0.4$ &$0.9$& $0.4$ &0.4 &$z_{3}$& $0.6$ &$0.4$& $0.6$& $0.5$&0.5\\
$z_{4}$ & $0.6$& $0.3$ &$0.6$& $0.8$ & 0.8&$z_{4}$& $0.6$ &$0.8$&$0.5$& $0.8$&0.8&$z_{4}$ & $0.8$& $0.3$ &$0.7$& $0.8$ &0.8 &$z_{4}$& $0.5$ &$0.3$& $0.5$& $0.8$&0.6\\
$z_{5}$ & $0.6$& $0.3$ &$0.6$& $0.8$ & 0.8&$z_{5}$& $0.6$ &$0.8$&$0.5$& $0.8$&0.8&$z_{5}$ & $0.8$& $0.3$ &$0.7$& $0.8$ &0.8 &$z_{5}$& $0.6$ &$0.3$& $0.6$& $0.8$&0.2\\
\hline
\end{tabular}
\end{table}

We compress the dynamic FRIS when adding a fuzzy relation. More concretely, we can
compress the updated FRIS by utilizing
$U_{1}/R_{1}$, $U_{1}/R_{2}$ and $U_{1}/R_{3}$ derived for the
original FRIS. The same approach can be
applied to the dynamic FRIS when
deleting a family of fuzzy relations.

\subsection{Emigration of fuzzy relations}

Suppose $S_{1}=(U_{1},\mathscr{R})$ is a fuzzy relation
information system. By deleting $R_{l}\in \mathscr{R}$, we get
$S_{2}=(U_{1}, \mathscr{R}-\{R_{l}\})$. There are three steps to
compress $S_{2}$ by utilizing the compression of $S_{1}$. First, we
obtain Table 10 by deleting $U_{1}/R_{l}$ shown in Table 2. Second,
we get $U_{1}/(\mathscr{R}-\{R_{l}\})$ based on
$\{U_{1}/R_{i}|1\leq i\leq l-1, l+1\leq i\leq m \}$ and define the
homomorphism $g$ as Example 3.12. Third, we obtain $S_{3}=(g(U_{1}),
g(\mathscr{R}-\{R_{l}\}))$. Furthermore, we can compress the dynamic fuzzy relation
information system when deleting a family of fuzzy relations.
\begin{algorithm}
	\label{algo:add}
	\SetKwData{Left}{left}
	\SetKwData{This}{this}
	\SetKwData{Up}{up}
	\SetKwFunction{Union}{Union}
	\SetKwInOut{Input}{input}
	\SetKwInOut{Output}{output}
	\SetKwBlock{Begin}{begin}{end}
	\SetKwIF{If}{ElseIf}{Else}{if}{then}{else if}{else}{endif}
	\SetKwComment{Comment}{//}{}
	\caption{Incremental algorithm for compressing FRIS under homomorphisms when removing an fuzzy relation}
	\Input{An fuzzy relation $R^+$ and
		$C_{\mathscr{R}}$. }
	\Output{The image system $(U_{2}^+, f^+(\mathscr{R}^+))$ and the partition $C_{f^+(\mathscr{R}^+)}$. The partition $C_{R_i^+}$ and $C_{\mathscr{R}^+}$.}
	
	\Begin{
		Compute $C_{R^+}$=$U_1/R^+$=$\{[x]_{R^+}\mid x \in U_1\}$.
		
		\BlankLine
		Compute $U_1/\mathscr{R}^+=\{[x]_{\mathscr{R}^+}\mid x \in U_1\}=C_{\mathscr{R}}\cap  C_{R^+}$.
		
		\BlankLine
		Denote $C_{\mathscr{R}^+}=U_1/\mathscr{R}^+=\{C_1, C_2, ..., C_s\}$
		
		\BlankLine
		Define $f^+: C_i \rightarrow y_i, 1 \leq i \leq s$
		, then $U_{2}^+=\{y_{1},y_{2},...,y_{s}\}$ and $C_{f^+(\mathscr{R}^+)}=\{\{y_1\}, \{y_2\}, ..., \{y_s\}\}$.
		
		\BlankLine
		For every $R_i$ in $\mathscr{R}^+$, compute $C_{f^+(R_{i})}$ from $C_{R_i}$ by replacing the $[x]_{R_i}$ with $y_1$ to $y_s$.
		
		\BlankLine
		Output $U_{2}^+$, $C_{f^+(R_i)}$, $C_{f^+(\mathscr{R}^+)}$,  $C_{R^+}$ and $C_{\mathscr{R}^+}$.\\
		
	}
\end{algorithm}

An example is employed to illustrate the process of compressing the
dynamic FRIS when deleting a fuzzy
relation.

\begin{example}\label{exam:del}
By deleting $R_{1}$ in $S_{1}$ shown in Table 3, we obtain
$S_{2}=(U_{1}, \mathscr{R}_{2})$, where
$\mathscr{R}_{2}=\{R_{2},R_{3}\}$. To compress $S_{2}$ based on the
compression of $S_{1}$, we get Table 11 by deleting $U_{1}/R_{1}$
from Table 4. Then we obtain $U_{1}/\mathscr{R}_{2}=\{\{x_{1},
x_{7}\},\{x_{2}, x_{6}\}, \{x_{3},x_{5}\},\{x_{4}, x_{8}\}\}$ and
define $g: U_{1}\rightarrow U_{2}$ as follows:
$$g(x_{1})=g(x_{7})= z_{1}, g(x_{2})=g(x_{6})
= z_{2}, g(x_{3})=g(x_{5})=z_{3}, g(x_{4})=g(x_{8})=z_{4},$$ where
$U_{2}=\{z_{1},z_{2},z_{3}\}$. Subsequently, $(U_{1},
\mathscr{R}-\{R_{1}\})$ can be compressed into $(U_{2},
\{g(R_{2}),g(R_{3})\})$ shown in Table 12.
\end{example}

\begin{table}[h!]
\caption{The partitions with respect to $R_{1}, R_{2}, R_{3}, R^+$
, $\mathscr{R}^+$ and $\mathscr{R}^-$.}
\tabcolsep0.13in
\label{tab:partitiondel}
\begin{tabular}{c c c c c c c}
\hline $U_{1}$ & $U_{1}/R_{1}$ & $U_{1}/R_{2}$ & $U_{1}/R_{3}$ & $U_{1}/R^+$ & $U_{1}/\mathscr{R}^+$ & $U_{1}/\mathscr{R}^-$\\
\hline
$x_{1}$ & $\{x_{1}, x_{3}, x_{5}, x_{7}\}$& $\{x_{1}, x_{7}\}$ & $\{x_{1}, x_{4}, x_{7}, x_{8}\}$&$\{x_{1}, x_{7}\}$ &$\{x_{1}, x_{3}, x_{5}, x_{7}\}$& $\{x_{1}, x_{7}\}$\\
$x_{2}$ & $\{x_{2}, x_{6}\}$& $\{x_{2}, x_{4}, x_{6}, x_{8}\}$ &$\{x_{2}, x_{6}\}$&$\{x_{2}, x_{6}\}$&$\{x_{2}, x_{6}\}$& $\{x_{2}, x_{6}\}$\\
$x_{3}$ & $\{x_{1}, x_{3}, x_{5}, x_{7}\}$& $\{x_{3},x_{5}\}$ &$\{x_{3}, x_{5}\}$&$\{x_{3}, x_{5}\}$&$\{x_{1}, x_{3}, x_{5}, x_{7}\}$& $\{x_{3}, x_{5}\}$\\
$x_{4}$ & $\{x_{4}, x_{8}\}$& $\{x_{2}, x_{4}, x_{6}, x_{8}\}$ &$\{x_{1}, x_{4}, x_{7}, x_{8}\}$&$\{x_{4}\}$&$\{x_{4}\}$& $\{x_{4}, x_{8}\}$\\
$x_{5}$ & $\{x_{1}, x_{3}, x_{5}, x_{7}\}$& $\{x_{3}, x_{5}\}$& $\{x_{3}, x_{5}\}$&$\{x_{3}, x_{5}\}$ & $\{x_{1}, x_{3}, x_{5}, x_{7}\}$&$\{x_{3}, x_{5}\}$\\
$x_{6}$ & $\{x_{2}, x_{6}\}$& $\{x_{2}, x_{4}, x_{6}, x_{8}\}$&$\{x_{2}, x_{6}\}$&$\{x_{2}, x_{6}\}$& $\{x_{2}, x_{6}\}$&$\{x_{2}, x_{6}\}$\\
$x_{7}$ & $\{x_{1}, x_{3}, x_{5}, x_{7}\}$& $\{x_{1}, x_{7}\}$ &$\{x_{1}, x_{4}, x_{7}, x_{8}\}$&$\{x_{1}, x_{7}\}$  &$\{x_{1}, x_{3}, x_{5}, x_{7}\}$& $\{x_{1}, x_{7}\}$\\
$x_{8}$ & $\{x_{4}, x_{8}\}$& $\{x_{2}, x_{4}, x_{6}, x_{8}\}$&$\{x_{1}, x_{4}, x_{7}, x_{8}\}$&$\{x_{8}\}$&$\{x_{8}\}$& $\{x_{4}, x_{8}\}$\\
\hline
\end{tabular}
\end{table}

We compress the dynamic FRIS when deleting a fuzzy relation. More concretely, there is no
need to compute $U_{1}/R_{2}$ and $U_{1}/R_{3}$ by utilizing the
results of the original information system, and the same approach
can be applied to the FRIS when
deleting a family of fuzzy relations. In addition, we can obtain the
results shown in Table 12 by deleting $g(R_{1})$ presented in Table
4 since $\{R_{2},R_{3}\}$ is a reduction of $(U_{1},\mathscr{R})$.

The computational complexity of
constructing $g$ is $\mathscr{O}((m-1)\ast n)$ with the incremental
algorithm. But the computational complexity is
$(m-1)\ast\mathscr{O}(n^{2})+\mathscr{O}((m-1)\ast n)$ without Table
2.


%
%

\subsection{Immigration of objects}

In this subsection, we present two methods for constructing the
homomorphisms between FRIS with
respect to immigration of objects.

Suppose $S_{1}=(U_{1}, \mathscr{R})$ is a fuzzy relation
information system, where $U_{1}=\{x_{1},x_{2},...,x_{n}\}$ and
$\mathscr{R}=\{R_{1}, R_{2},...,R_{m}\}$. By adding $\{x_{n+1},
x_{n+2},..., x_{n+t}\}$ into $U_{1}$, we obtain
$S_{2}=(U_{2},\mathscr{R}^{+})$, where
$U_{2}=\{x_{1},x_{2},...,x_{n+t}\}$ and
$\mathscr{R}^{+}=\{R^{+}_{1}, R^{+}_{2},...,R^{+}_{m}\}$.  By
cutting $R^{+}\in \mathscr{R}^{+}$ shown in Table 13 into $R$,
$R^{0}$ and $R^{\ast}$ shown in Tables 14, 15 and 16, respectively,
we illustrate the relationship between $R$ and $R^{+}$ exactly.

We introduce two concepts of consistent functions with respect to
two families of fuzzy relations for compressing dynamic fuzzy
relation information systems when adding objects.

\begin{definition}
Let $(U_{2}, \mathscr{R}^{+})$ be the updated fuzzy relation
information system of $(U_{1}, \mathscr{R})$, $f$ a mapping from
$U_{1}$ to $V_{1}$, $[x]_{f}=\{y|f(x)=f(y),y\in U_{1}\}$, and
$\mathscr{R}^{0}=\{R^{0}|R^{+}\in \mathscr{R}^{+}\}$. For
any $R^{0}\in \mathscr{R}^{0}_{1}$, $f$ is said to be a consistent
function with respect to $R^{0}$ if $[x]_{f}=[x]_{R^{0}}$ for any
$x\in U_{1}$, where $[x]_{R^{0}}=\{y|R^{0}(x,z)=R^{0}(y,z),y\in
U_{1}, z\in U_{2}-U_{1}\}$.
\end{definition}

\begin{definition}
Let $(U_{2}, \mathscr{R}^{+})$ be the updated fuzzy relation
information system of $(U_{1}, \mathscr{R})$, $f$ a mapping from
$U_{2}$ to $V_{1}$, $[x]_{f}=\{y|f(x)=f(y),y\in U_{2}-U_{1}\}$, and
$\mathscr{R}^{\ast}=\{R^{\ast}|R^{+}\in \mathscr{R}^{+}\}$.
For any $R^{\ast}\in \mathscr{R}^{\ast}_{1}$, $f$ is said to be a
consistent function with respect to $R^{\ast}$ if
$[x]_{f}=[x]_{R^{\ast}}$ for any $x\in U_{2}-U_{1}$, where
$[x]_{R^{\ast}}=\{y|R^{\ast}(x,z)=R^{\ast}(y,z),z\in U_{2},y\in
U_{2}-U_{1}\}$.
\end{definition}

Additionally, $f$ is said to be a consistent function with respect
to $\mathscr{R}^{0}$ (respectively, $\mathscr{R}^{\ast}$) if
$[x]_{f}=\bigcap\{[x]_{R^{\ast}}|R^{\ast}\in \mathscr{R}^{0}\}$
(respectively, $[x]_{f}=\bigcap\{[x]_{R^{\ast}}|R^{\ast}\in
\mathscr{R}^{\ast}\}$). For convenience, we denote $[x]_{f}$ as
$[x]_{\mathscr{R}^{0}}$ (respectively,
$[x]_{\mathscr{R}^{\ast}}$) if $f$ is a consistent function with
respect to $\mathscr{R}^{0}$ (respectively,
$\mathscr{R}^{\ast}$). Then we propose two approaches to
constructing homomorphisms between fuzzy relation information
systems.

{\bf{Approach 1:}} There are four steps to compress $S_{2}$ by
utilizing the compression of $S_{1}$. First, we obtain
$U_{2}/\Delta_{R^{+}}$ shown in Table 17, where $\Delta_{R^{+}}=\{R,
R^{0},R^{\ast}\}$. Concretely, we get $\{[x_{i}]_{R^{0}}|1\leq i\leq
n \}$ and $\{[x_{i}]_{R^{\ast}}|n+1\leq i\leq n+t \}$ of $\{x_{1},
x_{2},..., x_{n}\}$ and $\{x_{n+1}, x_{n+2},..., x_{n+t}\}$ based on
$R^{0}$ and $R^{\ast}$, respectively. Then we obtain
$U_{2}/\Delta_{R^{+}}=\{[x_{i}]_{\Delta_{R^{+}}}|x_{i}\in U_{2}\}$,
where $[x_{i}]_{\Delta_{R^{+}}}=[x_{i}]_{R}\cap [x_{i}]_{R^{0}}$
$(1\leq i\leq n)$ and $[x_{j}]_{\Delta_{R^{+}}}=[x_{j}]_{R^{\ast}}$
$(n+1\leq j\leq n+t)$. Second, we obtain
$U_{1}/\Delta=\{[x_{i}]_{\Delta}|x_{i}\in U_{2}\}=\{C_{j}|1\leq j
\leq N\}$ shown in Table 18, where
$\Delta=\{\Delta_{R^{+}_{i}}|R^{+}_{i}\in\mathscr{R}^{+}\}$ and
$[x_{i}]_{\Delta }=\bigcap_{R^{+}_{j}\in\mathscr{R}^{+}}
[x_{i}]_{\Delta_{R^{+}_{j}}}$. Third, we define $g$ as $g(x)=z_{i}$
for any $x\in C_{i}$ and get $S_{3}=(U_{3},
g(\mathscr{R}^{+}))$, where $U_{3}=\{z_{i}|1\leq i\leq N\}$.
Fourth, we obtain $S_{4}$ by compressing $S_{3}$ as $S_{1}$ shown in
Example 3.12. The computational complexity of constructing the
homomorphism is $ m\ast( \mathscr{O}(t\ast
n)+\mathscr{O}(t\ast(n+t))+\mathscr{O}(3\ast(n+t)))$ with the
incremental algorithm. But the computational complexity is
$\mathscr{O}(m\ast (n+t)\ast(n+t))$ without Table 2.

\begin{table}[htbp]\renewcommand{\arraystretch}{1.5}
\caption{The fuzzy relation $R^{+}$.}
 \tabcolsep0.145in
\begin{tabular}{c c c c c c c c c c c c c c c c c c c c c}
\hline $R^{+}$  & $x_{1}$ &$x_{2}$&.&.&.& $x_{n}$& $x_{n+1}$
&$.$&.&.& $x_{n+t}$\\
\hline
$x_{1}$ & $a_{11}$& $a_{12}$ &$.$&.&.& $a_{1n}$ & $a_{1(n+1)}$& $.$&.&. &$a_{1(n+t)}$\\
$x_{2}$ & $a_{21}$& $a_{22}$ &.&.&.& $a_{2n}$ & $a_{2(n+1)}$& .&.&. &$a_{2(n+t)}$\\
$.$ & $.$& $.$ &.&.&.& .&.&. & .& $.$ &$.$\\
$.$ & $.$& $.$ &.&.&.& $.$ & .&.&.& $.$ &$.$\\
$.$ & $.$& $.$ &.&.&.& $.$ & .&.&.& $.$ &$.$\\
$x_{n}$ & $a_{n1}$& $a_{n2}$ &.&.&.& $a_{nn}$ & $a_{n(n+1)}$& .&.&. &$a_{n(n+t)}$\\
$x_{n+1}$ & $a_{(n+1)1}$& $a_{(n+1)2}$ &.&.&.& $a_{(n+1)n}$ & $a_{(n+1)(n+1)}$& .&.&. &$a_{(n+1)(n+t)}$\\
$.$ & $.$& $.$ &.&.&.& .&.&. & $.$& $.$ &$.$\\
$.$ & $.$& $.$ &.&.&.& .&.&. & $.$& $.$ &$.$\\
$.$ & $.$& $.$ &.&.&.& $.$ & .&.&.& $.$ &$.$\\
$x_{n+t}$ & $a_{(n+t)1}$& $a_{(n+t)2}$ &.&.&.& $a_{(n+t)n}$ & $a_{(n+t)(n+1)}$& .&.&. &$a_{(n+t)(n+t)}$\\
\hline
\end{tabular}
\end{table}

\begin{table}[htbp]\renewcommand{\arraystretch}{1.5}
\caption{The part $R$ of $R^{+}$.}
 \tabcolsep0.3in
\begin{tabular}{c c c c c c c }
\hline $R$  & $x_{1}$ &$x_{2}$&.&.&.& $x_{n}$\\
\hline
$x_{1}$ & $a_{11}$& $a_{12}$ &$.$&.&.& $a_{1n}$ \\
$x_{2}$ & $a_{21}$& $a_{22}$ &.&.&.& $a_{2n}$\\
$.$ & $.$& $.$ &.& $.$ &$.$&$.$\\
$.$ & $.$& $.$ &.& $.$ &$.$&$.$\\
$.$ & $.$& $.$ &.& $.$ &$.$&$.$\\
$x_{n}$ & $a_{n1}$& $a_{n2}$ &.&.&.& $a_{nn}$ \\
\hline
\end{tabular}
\end{table}

\begin{table}[htbp]\renewcommand{\arraystretch}{1.5}
\caption{The part $R^{0}$ of $R^{+}$.}
 \tabcolsep0.35in
\begin{tabular}{c c c c c c c c c c c c c c c c c c c c c c}
\hline $R^{0}$ & $x_{n+1}$
&$x_{n+2}$&.&.&.& $x_{n+t}$\\
\hline
$x_{1}$ &$a_{1(n+1)}$&$a_{1(n+2)}$ &$.$&.&. &$a_{1(n+t)}$\\
$x_{2}$ & $a_{2(n+1)}$& $a_{2(n+2)}$&.&.&. &$a_{2(n+t)}$\\
$.$ & $.$& $.$ & .&.& $.$ &$.$\\
$.$ & $.$& $.$ &.&.& $.$ &$.$\\
$.$ & $.$& $.$ &.&.& $.$ &$.$\\
$x_{n}$ & $a_{n(n+1)}$&$a_{n(n+2)}$ &.&.&. &$a_{n(n+t)}$\\
\hline
\end{tabular}
\end{table}

\begin{table}[htbp]\renewcommand{\arraystretch}{1.5}
\caption{The part $R^{\ast}$ of $R^{+}$.}
 \tabcolsep0.145in
\begin{tabular}{c c c c c c c c c c c c c c c c c c c c c}
\hline $R^{\ast}$  & $x_{1}$ &$x_{2}$&.&.&.& $x_{n}$& $x_{n+1}$
&$.$&.&.& $x_{n+t}$\\
\hline
$x_{n+1}$ & $a_{(n+1)1}$& $a_{(n+1)2}$ &.&.&.& $a_{(n+1)n}$ & $a_{(n+1)(n+1)}$& .&.&. &$a_{(n+1)(n+t)}$\\
$x_{n+2}$ & $a_{(n+2)1}$& $a_{(n+2)2}$ &.&.&.& $a_{(n+2)n}$ & $a_{(n+2)(n+1)}$& .&.&. &$a_{(n+2)(n+t)}$\\
$.$ & $.$& $.$ &.&.&.& .&.&. & $.$& $.$ &$.$\\
$.$ & $.$& $.$ &.&.&.& .&.&. & $.$& $.$ &$.$\\
$.$ & $.$& $.$ &.&.&.& $.$ & .&.&.& $.$ &$.$\\
$x_{n+t}$ & $a_{(n+t)1}$& $a_{(n+t)2}$ &.&.&.& $a_{(n+t)n}$ & $a_{(n+t)(n+1)}$& .&.&. &$a_{(n+t)(n+t)}$\\
\hline
\end{tabular}
\end{table}

\begin{table}[htbp]\renewcommand{\arraystretch}{1.5}
\caption{The partitions with respect to $R$, $R^{0}$ and
$R^{\ast}$.}
 \tabcolsep0.4in
\begin{tabular}{c c c c c c c c}
\hline $U_{2}$  &$R$& $R^{0}$&$R^{\ast}$&$\Delta_{R^{+}}$\\
\hline
$x_{1}$ & $[x_{1}]_{R}$& $[x_{1}]_{R^{0}}$&$U_{1}$&$[x_{1}]_{\Delta_{R^{+}}}$\\
$x_{2}$ & $[x_{2}]_{R}$& $[x_{2}]_{R^{0}}$&$U_{1}$&$[x_{2}]_{\Delta_{R^{+}}}$\\
$.$     & $.$        & . &$ . $& .\\
$.$     & $.$        & . &$ . $&.\\
$.$     & $.$        & . &$ . $ &.\\
$x_{n}$ & $[x_{n}]_{R}$& $[x_{n}]_{R^{0}}$&$U_{1}$&$[x_{n}]_{\Delta_{R^{+}}}$\\
$x_{n+1}$ & $U_{2}-U_{1}$& $U_{2}-U_{1}$&$[x_{n+1}]_{R^{\ast}}$&$[x_{n+1}]_{\Delta_{R^{+}}}$\\
$.$     & $.$        & . &$ . $& .\\
$.$     & $.$        & . &$ . $&.\\
$.$     & $.$        & . &$ . $ &.\\
$x_{n+t}$ & $U_{2}-U_{1}$& $U_{2}-U_{1}$&$[x_{n+t}]_{R^{\ast}}$&$[x_{n+t}]_{\Delta_{R^{+}}}$  \\
\hline
\end{tabular}
\end{table}

\begin{table}[htbp]\renewcommand{\arraystretch}{1.5}
\caption{The partitions with respect to $\Delta_{R^{+}_{i}}$ $(1\leq
i\leq m )$ and $\Delta$.}
 \tabcolsep0.27in
\begin{tabular}{c c c c c c c c}
\hline $U_{1}$  &$\Delta_{R^{+}_{1}}$& $\Delta_{R^{+}_{2}}$&.&.&. &$\Delta_{R^{+}_{m}}$& $\Delta$\\
\hline
$x_{1}$ & $[x_{1}]_{\Delta_{R_{1}^{+}}}$& $[x_{1}]_{\Delta_{R_{2}^{+}}}$&.&.&. &$[x_{1}]_{\Delta_{R_{m}^{+}}}$ & $[ x_{1}]_{\Delta}$\\
$x_{2}$ & $[x_{2}]_{\Delta_{R_{1}^{+}}}$&$[x_{2}]_{\Delta_{R_{2}^{+}}}$&.&.&. &$[x_{2}]_{\Delta_{R_{m}^{+}}}$&$[x_{2}]_{\Delta}$\\
$.$     & $.$        & .&.&.&. &$ . $& .\\
$.$     & $.$        & .&.&.&. &$ . $&.\\
$.$     & $.$        & .&.&.&. &$ . $ &.\\
$x_{n}$ & $[x_{n}]_{\Delta_{R_{1}^{+}}}$& $[x_{n}]_{\Delta_{R_{2}^{+}}}$&.&.&. &$[x_{n}]_{\Delta_{R_{m}^{+}}}$& $[x_{n}]_{\Delta}$ \\
\hline
\end{tabular}
\end{table}

\begin{table}[htbp]\renewcommand{\arraystretch}{1.5}
\caption{The updated fuzzy relation information system
$(U_{2},\mathscr{R}^{+})$.}
 \tabcolsep0.092in
\begin{tabular}{c c c c c c c c c c c c c c c c c}
\hline $R^{+}_{1}$  & $x_{1}$ &$x_{2}$& $x_{3}$& $x_{4}$& $x_{5}$
&$x_{6}$& $x_{7}$& $x_{8}$& $x_{9}$& $x_{10}$& $R^{+}_{2}$& $x_{1}$ &$x_{2}$& $x_{3}$& $x_{4}$& $x_{5}$\\
\hline
$x_{1}$ & $0.7$& $0.4$ &$0.7$& $0.5$ & $0.7$& $0.4$ &$0.7$& $0.5$&$0.5$& $0.7$& $x_{1}$&$0.4$& $0.5$ &$0.7$& $0.5$& $0.7$\\
$x_{2}$ & $0.7$& $0.3$ &$0.7$& $0.8$ & $0.7$& $0.3$ &$0.7$& $0.8$&$0.8$& $0.7$& $x_{2}$&$0.6$& $0.8$ &$0.5$& $0.8$& $0.5$\\
$x_{3}$ & $0.7$& $0.4$ &$0.7$& $0.5$ & $0.7$& $0.4$ &$0.7$& $0.5$&$0.5$& $0.7$& $x_{3}$&$0.7$& $0.9$ &$0.2$& $0.9$& $0.2$\\
$x_{4}$ & $0.6$& $0.3$ &$0.6$& $0.8$ & $0.6$& $0.3$ &$0.6$& $0.8$&$0.8$& $0.6$& $x_{4}$&$0.6$& $0.8$ &$0.5$& $0.8$& $0.5$\\
$x_{5}$ & $0.7$& $0.4$ &$0.7$& $0.5$ & $0.7$& $0.4$ &$0.7$& $0.5$&$0.5$& $0.7$& $x_{5}$&$0.7$& $0.9$ &$0.2$& $0.9$& $0.2$\\
$x_{6}$ & $0.7$& $0.3$ &$0.7$& $0.8$ & $0.7$& $0.3$ &$0.7$& $0.8$&$0.8$& $0.7$& $x_{6}$&$0.6$& $0.8$ &$0.5$& $0.8$& $0.5$\\
$x_{7}$ & $0.7$& $0.4$ &$0.7$& $0.5$ & $0.7$& $0.4$ &$0.7$& $0.5$&$0.5$& $0.7$& $x_{7}$&$0.4$& $0.5$ &$0.7$& $0.5$& $0.7$\\
$x_{8}$ & $0.6$& $0.3$ &$0.6$& $0.8$ & $0.6$& $0.3$ &$0.6$& $0.8$&$0.8$& $0.6$& $x_{8}$&$0.6$& $0.8$ &$0.5$& $0.8$& $0.5$\\
$x_{9}$ & $0.6$& $0.3$ &$0.6$& $0.8$ & $0.6$& $0.3$ &$0.6$& $0.8$&$0.8$& $0.6$& $x_{9}$&$0.6$& $0.8$ &$0.5$& $0.8$& $0.5$\\
$x_{10}$ & $0.7$& $0.4$&$0.7$& $0.5$ & $0.7$& $0.4$ &$0.7$& $0.5$&$0.5$& $0.7$& $x_{10}$&$0.7$&$0.9$&$0.2$&$0.9$&$0.2$\\
\hline $R^{+}_{2}$&$x_{6}$& $x_{7}$ &$x_{8}$& $x_{9}$ &$x_{10}$& $R^{+}_{3}$& $x_{1}$ &$x_{2}$& $x_{3}$& $x_{4}$& $x_{5}$& $x_{6}$& $x_{7}$ &$x_{8}$& $x_{9}$ &$x_{10}$\\
\hline
$x_{1}$&$0.5$ &$0.4$& $0.5$&$0.5$& $0.7$&$x_{1}$ & $0.8$& $0.3$ &$0.7$& $0.8$ & $0.7$& $0.3$ &$0.8$& $0.8$&$0.8$& $0.7$\\
$x_{2}$&$0.8$ &$0.6$& $0.8$&$0.8$& $0.5$&$x_{2}$ & $0.7$& $0.2$ &$0.6$& $0.7$ & $0.6$& $0.2$ &$0.7$& $0.7$&$0.7$& $0.6$\\
$x_{3}$&$0.9$ &$0.7$& $0.9$&$0.9$& $0.2$&$x_{3}$ & $0.4$& $0.4$ &$0.9$& $0.4$ & $0.9$& $0.4$ &$0.4$& $0.4$&$0.4$& $0.9$\\
$x_{4}$&$0.8$ &$0.6$& $0.8$&$0.8$& $0.5$&$x_{4}$ & $0.8$& $0.3$ &$0.7$& $0.8$ & $0.7$& $0.3$ &$0.8$& $0.8$&$0.8$& $0.7$\\
$x_{5}$&$0.9$ &$0.7$& $0.9$&$0.9$& $0.2$&$x_{5}$ & $0.4$& $0.4$ &$0.9$& $0.4$ & $0.9$& $0.4$ &$0.4$& $0.4$&$0.4$& $0.9$\\
$x_{6}$&$0.8$ &$0.6$& $0.8$&$0.8$& $0.5$&$x_{6}$ & $0.7$& $0.2$ &$0.6$& $0.7$ & $0.6$& $0.2$ &$0.7$& $0.7$&$0.7$& $0.6$\\
$x_{7}$&$0.5$ &$0.4$& $0.5$&$0.5$& $0.7$&$x_{7}$ & $0.8$& $0.3$ &$0.7$& $0.8$ & $0.7$& $0.3$ &$0.8$& $0.8$&$0.8$& $0.7$\\
$x_{8}$&$0.8$ &$0.6$& $0.8$&$0.8$& $0.5$&$x_{8}$ & $0.8$& $0.3$ &$0.7$& $0.8$ & $0.7$& $0.3$ &$0.8$& $0.8$&$0.8$& $0.7$\\
$x_{9}$&$0.8$ &$0.6$& $0.8$&$0.8$& $0.5$&$x_{9}$ & $0.8$& $0.3$ &$0.7$& $0.8$ & $0.7$& $0.3$ &$0.8$& $0.8$&$0.8$& $0.7$\\
$x_{10}$&$0.9$ &$0.7$& $0.9$&$0.9$& $0.2$&$x_{10}$ & $0.4$& $0.4$ &$0.9$& $0.4$ & $0.9$& $0.4$ &$0.4$& $0.4$&$0.4$& $0.9$\\
\hline
\end{tabular}
\end{table}

We illustrate the process of compressing the dynamic fuzzy relation
information systems when adding an object set with the following
example.

\begin{example}
Tables 3 and 19 show the original fuzzy relation information system
$S_{1}$ and the updated fuzzy relation information system $S_{2}$,
respectively. First, we get $U_{2}/\Delta_{R^{+}_{1}}$ shown in
Table 20. Similarly, we obtain $U_{2}/\Delta_{R^{+}_{2}}$ and
$U_{2}/\Delta_{R^{+}_{3}}$. Second, we get $U_{2}/\Delta=\{\{x_{1},
x_{7}\},\{x_{2}, x_{6}\},\{x_{3},x_{5}\}, \{x_{4},x_{8}\},
\{x_{9}\},\{x_{10}\}\}$ and define $g$ as follows:
$g(x_{1})=g(x_{7})=z_{1}, g(x_{2})=g(x_{6})=z_{2},
g(x_{3})=g(x_{5})=z_{3}, g(x_{4})=g(x_{8})=z_{4}, g(x_{9})=z_{5},
g(x_{10})=z_{6}$. Thus we compress $S_{2}$ into
$S_{3}=(U_{3},g(\mathscr{R}^{+}))$ shown in Table 21, where
$U_{3}=\{z_{i}|1\leq i\leq 6\}$. Third, we compress $S_{3}$ as
$S_{1}$ shown in Example 3.12. Concretely, we define $h$ as follows:
$$h(z_{1})=w_{1}, h(z_{2})=w_{2}, h(z_{3})=h(z_{6})=w_{3},
h(z_{4})=h(z_{5})=w_{4},$$ and get $S_{4}$ shown in Table 22.
\end{example}

\begin{table}[htbp]\renewcommand{\arraystretch}{1.5}
\caption{The partitions with respect to $R_{1}$, $R^{0}_{1}$ and
$R^{\ast}_{1}$.}
 \tabcolsep0.33in
\begin{tabular}{c c c c c c c c}
\hline $U_{2}$  &$R_{1}$& $R^{0}_{1}$&$R^{\ast}_{1}$&$\Delta_{R^{+}_{1}}$\\
\hline
$x_{1}$     & $\{x_{1},x_{3},x_{5},x_{7}\}$  & $\{x_{1},x_{3},x_{5},x_{7}\}$  &$U_{1}$     &$\{x_{1},x_{3},x_{5},x_{7}\}$\\
$x_{2}$     & $\{x_{2},x_{6}\}$              & $\{x_{2},x_{6}\}$              &$U_{1}$     &$\{x_{2},x_{6}\}$\\
$x_{3}$     & $\{x_{1},x_{3},x_{5},x_{7}\}$  & $\{x_{1},x_{3},x_{5},x_{7}\}$  &$U_{1}$     &$\{x_{1},x_{3},x_{5},x_{7}\}$\\
$x_{4}$     & $\{x_{4},x_{8}\}$              & $\{x_{4},x_{8}\}$              &$U_{1}$     &$\{x_{4},x_{8}\}$\\
$x_{5}$     & $\{x_{1},x_{3},x_{5},x_{7}\}$  & $\{x_{1},x_{3},x_{5},x_{7}\}$  &$U_{1}$     &$\{x_{1},x_{3},x_{5},x_{7}\}$\\
$x_{6}$     & $\{x_{2},x_{6}\}$              & $\{x_{2},x_{6}\}$              &$U_{1}$     &$\{x_{2},x_{6}\}$\\
$x_{7}$     & $\{x_{1},x_{3},x_{5},x_{7}\}$  & $\{x_{1},x_{3},x_{5},x_{7}\}$  &$U_{1}$     &$\{x_{1},x_{3},x_{5},x_{7}\}$\\
$x_{8}$     & $\{x_{4},x_{8}\}$              & $\{x_{4},x_{8}\}$              &$U_{1}$     &$\{x_{4},x_{8}\}$\\
$x_{9}$     & $\{x_{9},x_{10}\}$                       & $\{x_{9},x_{10}\}$                        &$\{x_{9}\}$ &$\{x_{9}\}$\\
$x_{10}$    & $\{x_{9},x_{10}\}$                       & $\{x_{9},x_{10}\}$                        &$\{x_{10}\}$ &$\{x_{10}\}$  \\
\hline
\end{tabular}
\end{table}

\begin{table}[htbp]\renewcommand{\arraystretch}{1.5}
\caption{The fuzzy relation information system
$S_{3}=(U_{2},g(\mathscr{R}^{+}))$.}
 \tabcolsep0.045in
\begin{tabular}{c c c c c c c c c c c c c c c c c c c c c}
\hline $g(R^{+}_{1})$  & $z_{1}$ &$z_{2}$& $z_{3}$& $z_{4}$&
$z_{5}$& $z_{6}$& $g(R^{+}_{2})$ &$z_{1}$& $z_{2}$& $z_{3}$&
$z_{4}$& $z_{5}$& $z_{6}$& $g(R^{+}_{3})$ &$z_{1}$& $z_{2}$&
$z_{3}$& $z_{4}$& $z_{5}$&
$z_{6}$\\
\hline
$z_{1}$ & $0.7$& $0.4$ &$0.7$& $0.5$ &$0.5$& $0.7$& $z_{1}$& $0.4$ &$0.5$& $0.7$& $0.5$& $0.5$& $0.7$&$z_{1}$& $0.8$ &$0.3$& $0.7$& $0.8$& $0.8$& $0.7$\\
$z_{2}$ & $0.7$& $0.3$ &$0.7$& $0.8$ &$0.8$& $0.7$& $z_{2}$& $0.6$ &$0.8$& $0.5$& $0.8$& $0.8$& $0.5$&$z_{2}$& $0.7$ &$0.2$& $0.6$& $0.7$& $0.7$& $0.6$\\
$z_{3}$ & $0.7$& $0.4$ &$0.7$& $0.5$ &$0.5$& $0.7$& $z_{3}$& $0.7$ &$0.9$& $0.2$& $0.9$& $0.9$& $0.2$&$z_{3}$& $0.4$ &$0.4$& $0.9$& $0.4$& $0.4$& $0.9$\\
$z_{4}$ & $0.6$& $0.3$ &$0.6$& $0.8$ &$0.8$& $0.6$& $z_{4}$& $0.6$ &$0.8$& $0.5$& $0.8$& $0.8$& $0.5$&$z_{4}$& $0.8$ &$0.3$& $0.7$& $0.8$& $0.8$& $0.7$\\
$z_{5}$ & $0.6$& $0.3$ &$0.6$& $0.8$ &$0.8$& $0.6$& $z_{5}$& $0.6$ &$0.8$& $0.5$& $0.8$& $0.8$& $0.5$&$z_{5}$& $0.8$ &$0.3$& $0.7$& $0.8$& $0.8$& $0.7$\\
$z_{6}$ & $0.7$& $0.4$ &$0.7$& $0.5$ &$0.5$& $0.7$& $z_{6}$& $0.7$ &$0.9$& $0.2$& $0.9$& $0.9$& $0.2$&$z_{6}$& $0.4$ &$0.4$& $0.9$& $0.4$& $0.4$& $0.9$\\
\hline
\end{tabular}
\end{table}

\begin{table}[htbp]\renewcommand{\arraystretch}{1.5}
\caption{The fuzzy relation information system $S_{4}=(U_{3},h\circ
g(\mathscr{R}^{+}))$.}
 \tabcolsep0.08in
\begin{tabular}{c c c c c c c c c c c c c c c}
\hline $h\circ g(R^{+}_{1})$  & $w_{1}$ &$w_{2}$& $w_{3}$& $w_{4}$&
$h\circ g(R^{+}_{2})$
&$w_{1}$& $w_{2}$& $w_{3}$& $w_{4}$& $h\circ g(R^{+}_{3})$ &$w_{1}$& $w_{2}$& $w_{3}$& $w_{4}$\\
\hline
$w_{1}$ & $0.7$& $0.4$ &$0.7$& $0.5$ & $w_{1}$& $0.4$ &$0.5$& $0.7$& $0.5$&$w_{1}$& $0.8$ &$0.3$& $0.7$& $0.8$\\
$w_{2}$ & $0.7$& $0.3$ &$0.7$& $0.8$ & $w_{2}$& $0.6$ &$0.8$& $0.5$& $0.8$&$w_{2}$& $0.7$ &$0.2$& $0.6$& $0.7$\\
$w_{3}$ & $0.7$& $0.4$ &$0.7$& $0.5$ & $w_{3}$& $0.7$ &$0.9$& $0.2$& $0.9$&$w_{3}$& $0.4$ &$0.4$& $0.9$& $0.4$\\
$w_{4}$ & $0.6$& $0.3$ &$0.6$& $0.8$ & $w_{4}$& $0.6$ &$0.8$& $0.5$& $0.8$&$w_{4}$& $0.8$ &$0.3$& $0.7$& $0.8$\\
\hline
\end{tabular}
\end{table}

{\bf{Approach 2:}} There are four steps to compress $S_{2}$ by
utilizing the compression of $S_{1}$. First, we obtain
$U_{1}/\mathscr{R}^{0}=\{[x]_{\mathscr{R}^{0}}|x\in U_{1}\}$
and
$(U_{2}-U_{1})/\mathscr{R}^{\ast}=\{[x]_{\mathscr{R}^{\ast}}|x\in
U_{2}-U_{1}\}$ shown in Tables 23 and 24, respectively. Concretely,
$[x]_{\mathscr{R}^{0}}=\bigcap_{R^{0}\in
\mathscr{R}^{0}}[x]_{R^{0}}$ and
$[x]_{\mathscr{R}^{\ast}}=\bigcap_{R^{\ast}\in
\mathscr{R}^{\ast}}[x]_{R^{\ast}}$. Second, we derive
$U_{2}/\Delta_{\mathscr{R}^{+}}=\{[x]_{\Delta_{\mathscr{R}^{+}}}|x\in
U_{2}\}=\{C_{i}|1\leq i\leq N\}$ shown in Table 25, where
$\Delta_{\mathscr{R}^{+}}=\{\mathscr{R},\mathscr{R}^{0},
\mathscr{R}^{\ast}\}$. Concretely, we have that
$[x]_{\Delta_{\mathscr{R}^{+}}}=[x]_{\mathscr{R}}\cap
[x]_{\mathscr{R}^{0}}$ and $[x]_{\Delta_{\mathscr{R}^{+}}}=
[x]_{\mathscr{R}^{\ast}}$ for $x\in U_{1}$ and $x\in
U_{2}-U_{1}$, respectively. Third, we define $g$ as $g(x)=z_{i}$ for
any $x\in C_{i}$ and get $S_{3}=(U_{3}, g(\mathscr{R}^{+}))$,
where $U_{3}=\{z_{i}|1\leq i\leq N\}$. Fourth, we obtain $S_{4}$ by
compressing $S_{3}$ as $S_{1}$ shown in Example 3.12. The
computational complexity of constructing the homomorphism is $
m\ast( \mathscr{O}(t\ast
n)+\mathscr{O}(t\ast(n+t)))+2\ast\mathscr{O}(m\ast(n+t))+\mathscr{O}(3\ast(n+t))$
with the incremental algorithm. But the computational complexity is
$\mathscr{O}(m\ast (n+t)\ast(n+t))$ without Table 2.

\begin{table}[htbp]\renewcommand{\arraystretch}{1.5}
\caption{The partitions with respect to $R^{0}\in
\mathscr{R}^{0}$ and $\mathscr{R}^{0}$.}
 \tabcolsep0.245in
\begin{tabular}{c c c c c c c c c c c}
\hline $U_{2}$  &$R_{1}^{0}$& $R_{2}^{0}$&.&.&.&$R_{m}^{0}$&$\mathscr{R}^{0}$\\
\hline
$x_{1}$ & $[x_{1}]_{R_{1}^{0}}$& $[x_{1}]_{R_{2}^{0}}$&.&.&.&$[x_{1}]_{R_{m}^{0}}$&$[x_{1}]_{\mathscr{R}^{0}}$\\
$x_{2}$ & $[x_{2}]_{R_{1}^{0}}$& $[x_{2}]_{R_{2}^{0}}$&.&.&.&$[x_{2}]_{R_{m}^{0}}$&$[x_{2}]_{\mathscr{R}^{0}}$\\
$.$     & $.$        & .&.&.& .&$ . $& .\\
$.$     & $.$        & .&.&.& .&$ . $& .\\
$.$     & $.$        & .&.&.& .&$ . $& .\\
$x_{n}$ & $[x_{n}]_{R_{1}^{0}}$& $[x_{n}]_{R_{2}^{0}}$&.&.&.&$[x_{n}]_{R_{m}^{0}}$&$[x_{n}]_{\mathscr{R}^{0}}$\\
$x_{n+1}$ & $U_{2}-U_{1}$& $U_{2}-U_{1}$&.&.&.&$U_{2}-U_{1}$&$U_{2}-U_{1}$\\
$.$     & $.$        & .&.&.& .&$ . $& .\\
$.$     & $.$        & .&.&.& .&$ . $& .\\
$.$     & $.$        & .&.&.& .&$ . $& .\\
$x_{n+t}$ & $U_{2}-U_{1}$& $U_{2}-U_{1}$&.&.&.&$U_{2}-U_{1}$&$U_{2}-U_{1}$ \\
\hline
\end{tabular}
\end{table}

\begin{table}[htbp]\renewcommand{\arraystretch}{1.5}
\caption{The partition with respect to $R^{\ast}\in
\mathscr{R}^{\ast}$.}
 \tabcolsep0.245in
\begin{tabular}{c c c c c c c c c c c}
\hline $U_{2}$  &$R_{1}^{\ast}$& $R_{2}^{\ast}$&.&.&.&$R_{m}^{\ast}$&$\mathscr{R}^{\ast}$\\
\hline
$x_{1}$ & $U_{1}$& $U_{1}$&.&.&.&$U_{1}$&$U_{1}$\\
$x_{2}$ & $U_{1}$& $U_{1}$&.&.&.&$U_{1}$&$U_{1}$\\
$.$     & $.$        & .&.&.& .&$ . $& .\\
$.$     & $.$        & .&.&.& .&$ . $& .\\
$.$     & $.$        & .&.&.& .&$ . $& .\\
$x_{n}$ & $U_{1}$& $U_{1}$&.&.&.&$U_{1}$&$U_{1}$\\
$x_{n+1}$ & $[x_{n+1}]_{R_{1}^{\ast}}$& $[x_{n+1}]_{R_{2}^{\ast}}$&.&.&.&$[x_{n+1}]_{R_{m}^{\ast}}$&$[x_{n+1}]_{\mathscr{R}^{\ast}}$\\
$.$     & $.$        & .&.&.& .&$ . $& .\\
$.$     & $.$        & .&.&.& .&$ . $& .\\
$.$     & $.$        & .&.&.& .&$ . $& .\\
$x_{n+t}$ & $[x_{n+t}]_{R_{1}^{\ast}}$& $[x_{n+t}]_{R_{2}^{\ast}}$&.&.&.&$[x_{n+t}]_{R_{m}^{\ast}}$&$[x_{n+t}]_{\mathscr{R}^{\ast}}$ \\
\hline
\end{tabular}
\end{table}

\begin{table}[htbp]\renewcommand{\arraystretch}{1.5}
\caption{The partitions with respect to $\mathscr{R}$,
$\mathscr{R}^{0}$, $\mathscr{R}^{\ast}$ and
$\Delta_{\mathscr{R}^{+}}$.}
 \tabcolsep0.38in
\begin{tabular}{c c c c c c c c}
\hline $U_{2}$  &$\mathscr{R}$& $\mathscr{R}^{0}$&$\mathscr{R}^{\ast}$&$\Delta_{\mathscr{R}^{+}}$\\
\hline
$x_{1}$ & $[x_{1}]_{\mathscr{R}}$& $[x_{1}]_{\mathscr{R}^{0}_{1}}$&$U_{1}$&$[x_{1}]_{\Delta_{\mathscr{R}^{+}}}$\\
$x_{2}$ & $[x_{2}]_{\mathscr{R}}$& $[x_{2}]_{\mathscr{R}^{0}_{1}}$&$U_{1}$&$[x_{2}]_{\Delta_{\mathscr{R}^{+}}}$\\
$.$     & $.$        & . &$ . $& .\\
$.$     & $.$        & . &$ . $&.\\
$.$     & $.$        & . &$ . $ &.\\
$x_{n}$ & $[x_{n}]_{\mathscr{R}}$& $[x_{n}]_{\mathscr{R}^{0}_{1}}$&$U_{1}$&$[x_{n}]_{\Delta_{\mathscr{R}^{+}}}$\\
$x_{n+1}$ & $U_{2}-U_{1}$& $U_{2}-U_{1}$&$[x_{n+1}]_{\mathscr{R}^{\ast}_{1}}$&$[x_{n+1}]_{\Delta_{\mathscr{R}^{+}}}$\\
$.$     & $.$        & . &$ . $& .\\
$.$     & $.$        & . &$ . $&.\\
$.$     & $.$        & . &$ . $ &.\\
$x_{n+t}$ & $U_{2}-U_{1}$& $U_{2}-U_{1}$&$[x_{n+t}]_{\mathscr{R}^{\ast}_{1}}$&$[x_{n+t}]_{\Delta_{\mathscr{R}^{+}}}$ \\
\hline
\end{tabular}
\end{table}

\begin{table}[htbp]\renewcommand{\arraystretch}{1.5}
\caption{The partitions with respect to $R_{1}^{0}$, $R_{2}^{0}$ and
$R_{3}^{0}$.}
 \tabcolsep0.3in
\begin{tabular}{c c c c c c c c}
\hline $U_{2}$  &$R_{1}^{0}$& $R_{2}^{0}$&$R_{3}^{0}$&$\mathscr{R}^{0}$\\
\hline
$x_{1}$ & $\{x_{1},x_{3},x_{5},x_{7}\}$        &$\{x_{1},x_{7}\}$              &$\{x_{1},x_{4},x_{7},x_{8}\}$    &$\{x_{1},x_{7}\}$\\
$x_{2}$ & $\{x_{2},x_{4},x_{6},x_{8}\}$        &$\{x_{2},x_{4},x_{6},x_{8}\}$  &$\{x_{2},x_{6}\}$                &$\{x_{2},x_{6}\}$\\
$x_{3}$ & $\{x_{1},x_{3},x_{5},x_{7}\}$        &$\{x_{3},x_{5}\}$              &$\{x_{3},x_{5}\}$                &$\{x_{3},x_{5}\}$\\
$x_{4}$ & $\{x_{2},x_{4},x_{6},x_{8}\}$        &$\{x_{2},x_{4},x_{6},x_{8}\}$  &$\{x_{1},x_{4},x_{7},x_{8}\}$    &$\{x_{4},x_{8}\}$\\
$x_{5}$ & $\{x_{1},x_{3},x_{5},x_{7}\}$        &$\{x_{3},x_{5}\}$              &$\{x_{3},x_{5}\}$                &$\{x_{3},x_{5}\}$\\
$x_{6}$ & $\{x_{2},x_{4},x_{6},x_{8}\}$        &$\{x_{2},x_{4},x_{6},x_{8}\}$  &$\{x_{2},x_{6}\}$                &$\{x_{2},x_{6}\}$\\
$x_{7}$ & $\{x_{1},x_{3},x_{5},x_{7}\}$        &$\{x_{1},x_{7}\}$              &$\{x_{1},x_{4},x_{7},x_{8}\}$    &$\{x_{1},x_{7}\}$\\
$x_{8}$ & $\{x_{2},x_{4},x_{6},x_{8}\}$        &$\{x_{2},x_{4},x_{6},x_{8}\}$  &$\{x_{1},x_{4},x_{7},x_{8}\}$    &$\{x_{4},x_{8}\}$\\
$x_{9}$ & $\{x_{9},x_{10}\}$                   &$\{x_{9},x_{10}\}$             &$\{x_{9},x_{10}\}$               &$\{x_{9},x_{10}\}$\\
$x_{10}$& $\{x_{9},x_{10}\}$                   &$\{x_{9},x_{10}\}$             &$\{x_{9},x_{10}\}$               &$\{x_{9},x_{10}\}$\\
\hline
\end{tabular}
\end{table}

\begin{table}[htbp]\renewcommand{\arraystretch}{1.5}
\caption{The partitions with respect to $R_{1}^{\ast}$,
$R_{2}^{\ast}$ and $R_{3}^{\ast}$.}
 \tabcolsep0.28in
\begin{tabular}{c c c c c c c c}
\hline $U_{2}$  &$R_{1}^{\ast}$& $R_{2}^{\ast}$&$R_{3}^{\ast}$&$\mathscr{R}^{\ast}$\\
\hline
$x_{1}$ & $\{x_{1},x_{2},...,x_{8}\}$        &$\{x_{1},x_{2},...,x_{8}\}$    &$\{x_{1},x_{2},...,x_{8}\}$    &$\{x_{1},x_{2},...,x_{8}\}$\\
$x_{2}$ & $\{x_{1},x_{2},...,x_{8}\}$        &$\{x_{1},x_{2},...,x_{8}\}$    &$\{x_{1},x_{2},...,x_{8}\}$    &$\{x_{1},x_{2},...,x_{8}\}$\\
$x_{3}$ & $\{x_{1},x_{2},...,x_{8}\}$        &$\{x_{1},x_{2},...,x_{8}\}$    &$\{x_{1},x_{2},...,x_{8}\}$    &$\{x_{1},x_{2},...,x_{8}\}$\\
$x_{4}$ & $\{x_{1},x_{2},...,x_{8}\}$        &$\{x_{1},x_{2},...,x_{8}\}$    &$\{x_{1},x_{2},...,x_{8}\}$    &$\{x_{1},x_{2},...,x_{8}\}$\\
$x_{5}$ & $\{x_{1},x_{2},...,x_{8}\}$        &$\{x_{1},x_{2},...,x_{8}\}$    &$\{x_{1},x_{2},...,x_{8}\}$    &$\{x_{1},x_{2},...,x_{8}\}$\\
$x_{6}$ & $\{x_{1},x_{2},...,x_{8}\}$        &$\{x_{1},x_{2},...,x_{8}\}$    &$\{x_{1},x_{2},...,x_{8}\}$    &$\{x_{1},x_{2},...,x_{8}\}$\\
$x_{7}$ & $\{x_{1},x_{2},...,x_{8}\}$        &$\{x_{1},x_{2},...,x_{8}\}$    &$\{x_{1},x_{2},...,x_{8}\}$    &$\{x_{1},x_{2},...,x_{8}\}$\\
$x_{8}$ & $\{x_{1},x_{2},...,x_{8}\}$        &$\{x_{1},x_{2},...,x_{8}\}$    &$\{x_{1},x_{2},...,x_{8}\}$    &$\{x_{1},x_{2},...,x_{8}\}$\\
$x_{9}$ & $\{x_{9}\}$                        &$\{x_{9}\}$                    &$\{x_{9}\}$                    &$\{x_{9}\}$\\
$x_{10}$& $\{x_{10}\}$                       &$\{x_{10}\}$                   &$\{x_{10}\}$                    &$\{x_{10}\}$\\
\hline
\end{tabular}
\end{table}

\begin{table}[htbp]\renewcommand{\arraystretch}{1.5}
\caption{The partitions with respect to $\mathscr{R}$,
$\mathscr{R}^{0}$ and $\mathscr{R}^{\ast}$.}
 \tabcolsep0.385in
\begin{tabular}{c c c c c c c c}
\hline $U_{2}$  &$\mathscr{R}$& $\mathscr{R}^{0}$&$\mathscr{R}^{\ast}$&$\Delta_{\mathscr{R}^{+}}$\\
\hline
$x_{1}$ & $\{x_{1},x_{7}\}$        &$\{x_{1},x_{7}\}$    &$\{x_{1},x_{2},...,x_{8}\}$    &$\{x_{1},x_{7}\}$\\
$x_{2}$ & $\{x_{2},x_{6}\}$        &$\{x_{2},x_{6}\}$    &$\{x_{1},x_{2},...,x_{8}\}$    &$\{x_{2},x_{6}\}$\\
$x_{3}$ & $\{x_{3},x_{5}\}$        &$\{x_{3},x_{5}\}$    &$\{x_{1},x_{2},...,x_{8}\}$    &$\{x_{3},x_{5}\}$\\
$x_{4}$ & $\{x_{4},x_{8}\}$        &$\{x_{4},x_{8}\}$    &$\{x_{1},x_{2},...,x_{8}\}$    &$\{x_{4},x_{8}\}$\\
$x_{5}$ & $\{x_{3},x_{5}\}$        &$\{x_{3},x_{5}\}$    &$\{x_{1},x_{2},...,x_{8}\}$    &$\{x_{3},x_{5}\}$\\
$x_{6}$ & $\{x_{2},x_{6}\}$        &$\{x_{2},x_{6}\}$    &$\{x_{1},x_{2},...,x_{8}\}$    &$\{x_{2},x_{6}\}$\\
$x_{7}$ & $\{x_{1},x_{7}\}$        &$\{x_{1},x_{7}\}$    &$\{x_{1},x_{2},...,x_{8}\}$    &$\{x_{1},x_{7}\}$\\
$x_{8}$ & $\{x_{4},x_{8}\}$        &$\{x_{4},x_{8}\}$    &$\{x_{1},x_{2},...,x_{8}\}$    &$\{x_{4},x_{8}\}$\\
$x_{9}$ & $\{x_{9},x_{10}\}$       &$\{x_{9},x_{10}\}$   &$\{x_{9}\}$                    &$\{x_{9}\}$\\
$x_{10}$& $\{x_{9},x_{10}\}$       &$\{x_{9},x_{10}\}$   &$\{x_{10}\}$                    &$\{x_{10}\}$\\
\hline
\end{tabular}
\end{table}

\begin{example} (Continuation of Example 4.5)
Tables 3 and 19 show the original fuzzy relation information system
$S_{1}$ and the updated fuzzy relation information system $S_{2}$,
respectively. First, we get $U_{2}/\mathscr{R}^{0}_{1}$ and
$U_{2}/\mathscr{R}^{\ast}_{1}$ shown in Tables 26 and 27,
respectively. Second, we get
$U_{2}/\Delta_{\mathscr{R}^{+}}$=$\{\{x_{1}, x_{7}\}$, $\{x_{2}$, $
x_{6}\}$, $\{x_{3},x_{5}\}$, $ \{x_{4},x_{8}\}$, $ \{x_{9}\}$, $\{x_{10}\}\}$
shown in Table 28 and define $g$ as follows:
$g(x_{1})=g(x_{7})=z_{1}, g(x_{2})=g(x_{6})=z_{2},
g(x_{3})=g(x_{5})=z_{3}, g(x_{4})=g(x_{8})=z_{4}, g(x_{9})=z_{5},
g(x_{10})=z_{6}$. Thus we compress $S_{2}$ into
$S_{3}=(U_{3},g(\mathscr{R}^{+}))$ shown in Table 21, where
$U_{3}=\{z_{i}|1\leq i\leq 6\}$. Third, we compress $S_{3}$ as
$S_{1}$ shown in Example 3.12. Concretely, we define $h$ as follows:
$$h(z_{1})=w_{1}, h(z_{2})=w_{2}, h(z_{3})=h(z_{6})=w_{3},
h(z_{4})=h(z_{5})=w_{4},$$ and get $S_{4}$ shown in Table 22.
\end{example}

It is obvious that the results are the same as that in Example 4.5.
Actually, the difference between Examples 4.5 and 4.6 is the
approach to computing the partition of the universe for constructing
homomorphisms between FRIS.

\subsection{Emigration of objects}

Suppose $S_{1}=(U_{1}, \mathscr{R})$ is a fuzzy relation
information system, we have obtained
$U_{1}/\mathscr{R}=\{[x]_{\mathscr{R}}|x\in U_{1}\}$ and
compressed $S_{1}$ to $S_{2}=(U_{2}, \mathscr{R}_{2})$ under the
condition of the homomorphism $f$. By deleting $\{x_{l+1},
x_{l+2},..., x_{n}\}$ in $U_{1}$, we obtain
$S_{3}=(U_{3},\mathscr{R}^{-})$, where
$U_{3}=\{x_{1},x_{2},...,x_{l}\}$ and
$\mathscr{R}^{-}=\{R^{-}_{1}, R^{-}_{2},...,R^{-}_{m}\}$. By
cutting $R$ shown in Table 14 into three parts: $R^{-}$, $R^{0-}$
and $R^{\ast-}$ shown in Tables 29, 30 and 31, respectively, we
illustrate the relationship between $R\in \mathscr{R}$ and
$R^{-}\in \mathscr{R}^{-}$. Furthermore, we get
$S_{4}=(U_{4},\mathscr{R}^{\ast-}_{1})$, where $U_{4}=\{x_{l+1},
x_{l+2},..., x_{n}\}$ and $\mathscr{R}^{\ast-}_{1}=\{R^{\ast-}|R\in
\mathscr{R}\}$.

There are three steps to compress $S_{3}=(U_{3},
\mathscr{R}^{-})$ based on $S_{2}$. First, as Example 3.12, we
obtain
$U_{4}/\mathscr{R}^{\ast-}_{1}=\{[x]_{\mathscr{R}^{\ast-}_{1}}|x\in
U_{4}\}$. It is obvious that $[x]_{\mathscr{R}^{\ast-}_{1}}\subseteq
[x]_{\mathscr{R}}$ for any $x\in U_{4}$. Second, we cancel the
object $f(x)$ in $U_{2}$ if
$[x]_{\mathscr{R}^{\ast-}_{1}}=[x]_{\mathscr{R}}$ and keep
$f(x)$ in $U_{2}$ if
$[x]_{\mathscr{R}^{\ast-}_{1}}\neq[x]_{\mathscr{R}}$. Then, we
obtain $S_{5}=(U_{5}, \mathscr{R}_{5})$, where $U_{5}=f(U_{3})$ and
$\mathscr{R}_{5}=f(\mathscr{R}^{-})$. Third, we get $S_{6}$ by
compressing $S_{5}$ as $S_{1}$ shown in Example 3.12. The
computational complexity of constructing the homomorphism is $
(n-l)\ast( \mathscr{O}(n\ast
(n-l))+m\ast\mathscr{O}(|U_{5}|^{2})+\mathscr{O}(m\ast|U_{5}|)$ with
the incremental algorithm. But the computational complexity is
$m\ast\mathscr{O}(l^{2})$ without Table 2.

\begin{table}[htbp]\renewcommand{\arraystretch}{1.5}
\caption{The part $R^{-}$ of $R$.}
 \tabcolsep0.3in
\begin{tabular}{c c c c c c c }
\hline $R^{-}$  & $x_{1}$ &$x_{2}$&.&.&.& $x_{l}$\\
\hline
$x_{1}$ & $a_{11}$& $a_{12}$ &$.$&.&.& $a_{1l}$ \\
$x_{2}$ & $a_{21}$& $a_{22}$ &.&.&.& $a_{2l}$\\
$.$ & $.$& $.$ &.& $.$ &$.$&$.$\\
$.$ & $.$& $.$ &.& $.$ &$.$&$.$\\
$.$ & $.$& $.$ &.& $.$ &$.$&$.$\\
$x_{l}$ & $a_{l1}$& $a_{l2}$ &.&.&.& $a_{ll}$ \\
\hline
\end{tabular}
\end{table}

\begin{table}[htbp]\renewcommand{\arraystretch}{1.5}
\caption{The part $R^{0-}$ of $R$.}
 \tabcolsep0.36in
\begin{tabular}{c c c c c c c }
\hline $R^{0-}$  & $x_{l+1}$ &$x_{l+2}$&.&.&.& $x_{n}$\\
\hline
$x_{1}$ & $a_{1(l+1)}$& $a_{1(l+2)}$ &$.$&.&.& $a_{nn}$ \\
$x_{2}$ & $a_{2(l+1)}$& $a_{2(l+2)}$ &.&.&.& $a_{2n}$\\
$.$ & $.$& $.$ &.& $.$ &$.$&$.$\\
$.$ & $.$& $.$ &.& $.$ &$.$&$.$\\
$.$ & $.$& $.$ &.& $.$ &$.$&$.$\\
$x_{l}$ & $a_{l(l+1)}$& $a_{l(l+2)}$ &.&.&.& $a_{ln}$ \\
\hline
\end{tabular}
\end{table}

\begin{table}[htbp]\renewcommand{\arraystretch}{1.5}
\caption{The part $R^{\ast-}$ of $R$.}
 \tabcolsep0.35in
\begin{tabular}{c c c c c c c }
\hline $R^{\ast-}$  & $x_{1}$ &$x_{2}$&.&.&.& $x_{n}$\\
\hline
$x_{l+1}$ & $a_{(l+1)1}$& $a_{(l+1)2}$ &$.$&.&.& $a_{(l+1)n}$ \\
$x_{l+2}$ & $a_{(l+2)1}$& $a_{(l+2)2}$ &.&.&.& $a_{(l+2)n}$\\
$.$ & $.$& $.$ &.& $.$ &$.$&$.$\\
$.$ & $.$& $.$ &.& $.$ &$.$&$.$\\
$.$ & $.$& $.$ &.& $.$ &$.$&$.$\\
$x_{n}$ & $a_{n1}$& $a_{n2}$ &.&.&.& $a_{nn}$ \\
\hline
\end{tabular}
\end{table}

We employ an example to show the process of compressing the dynamic
fuzzy relation information system when deleting an object set.

\begin{example} We take
$S_{1}=(U_{1},\mathscr{R})$ shown in Table 3. By deleting
objects $\{x_{1},x_{7},x_{8}\}$, we obtain $S_{3}=(U_{3},
\mathscr{R}^{-})$ and $S_{4}=(U_{4}, \mathscr{R}^{\ast-}_{1})$
shown in Table 32, where $U_{3}=\{x_{2},x_{3},x_{4},x_{5},x_{6}\}$
and $U_{4}=\{x_{1},x_{7},x_{8}\}$. As Example 4.6, we have that
$[x_{1}]_{\mathscr{R}^{\ast-}_{1}}=[x_{7}]_{\mathscr{R}^{\ast-}_{1}}=\{x_{1},x_{7}\}$
and $[x_{8}]_{\mathscr{R}^{\ast-}_{1}}=\{x_{8}\}$. Obviously,
$[x_{1}]_{\mathscr{R}^{\ast-}_{1}}=[x_{7}]_{\mathscr{R}^{\ast-}_{1}}=[x_{1}]_{\mathscr{R}}=[x_{7}]_{\mathscr{R}}$
and
$[x_{8}]_{\mathscr{R}^{\ast-}_{1}}\neq[x_{8}]_{\mathscr{R}}$.
Thus we delete $f(x_{1})$ in Table 5 and obtain $S_{5}$ shown in
Table 33. Finally, we get $S_{6}$ by compressing $S_{5}$ as $S_{1}$
shown in Example 3.12. We observe that $S_{6}=S_{5}$ in this
example.

\begin{table}[htbp]\renewcommand{\arraystretch}{1.5}
\caption{The fuzzy relation information system
$S_{4}=(U_{4},\mathscr{R}^{\ast-}_{1})$.}
 \tabcolsep0.021in
\begin{tabular}{c c c c c c c c c c c c c c c c c c c c c c c c c c c}
\hline $R^{\ast-}_{1}$  & $x_{1}$ &$x_{2}$& $x_{3}$& $x_{4}$&
$x_{5}$
&$x_{6}$& $x_{7}$& $x_{8}$& $R^{\ast-}_{2}$& $x_{1}$ &$x_{2}$& $x_{3}$& $x_{4}$& $x_{5}$& $x_{6}$& $x_{7}$ &$x_{8}$& $R^{\ast-}_{3}$& $x_{1}$ &$x_{2}$& $x_{3}$& $x_{4}$& $x_{5}$& $x_{6}$& $x_{7}$ &$x_{8}$\\
\hline
$x_{1}$ & $0.7$& $0.4$ &$0.7$& $0.5$ & $0.7$& $0.4$ &$0.7$& $0.5$& $x_{1}$&$0.4$& $0.5$ &$0.7$& $0.5$& $0.7$& $0.5$ &$0.4$& $0.5$&$x_{1}$ & $0.8$& $0.3$ &$0.7$& $0.8$ & $0.7$& $0.3$ &$0.8$& $0.8$\\
$x_{7}$ & $0.7$& $0.4$ &$0.7$& $0.5$ & $0.7$& $0.4$ &$0.7$& $0.5$& $x_{7}$&$0.4$& $0.5$ &$0.7$& $0.5$& $0.7$& $0.5$ &$0.4$& $0.5$&$x_{7}$ & $0.8$& $0.3$ &$0.7$& $0.8$ & $0.7$& $0.3$ &$0.8$& $0.8$\\
$x_{8}$ & $0.6$& $0.3$ &$0.6$& $0.8$ & $0.6$& $0.3$ &$0.6$& $0.8$& $x_{8}$&$0.6$& $0.8$ &$0.5$& $0.8$& $0.5$& $0.8$ &$0.6$& $0.8$&$x_{8}$ & $0.8$& $0.3$ &$0.7$& $0.8$ & $0.7$& $0.3$ &$0.8$& $0.8$\\
\hline
\end{tabular}
\end{table}

\begin{table}[htbp]\renewcommand{\arraystretch}{1.5}
\caption{The fuzzy relation information system
$S_{5}=(f(U_{3}),f(\mathscr{R}^{-}))$.}
 \tabcolsep0.14in
\begin{tabular}{c c c c c c c c c c c c c c c}
\hline $f(R^{-}_{1})$  & $y_{2}$& $y_{3}$& $y_{4}$& $f(R^{-}_{2})$
& $y_{2}$& $y_{3}$& $y_{4}$& $f(R^{-}_{3})$ & $y_{2}$& $y_{3}$& $y_{4}$\\
\hline
$y_{2}$ & $0.3$ &$0.7$& $0.8$ & $y_{2}$ &$0.8$& $0.5$& $0.8$&$y_{2}$ &$0.2$& $0.6$& $0.7$\\
$y_{3}$ & $0.4$ &$0.7$& $0.5$ & $y_{3}$ &$0.9$& $0.2$& $0.9$&$y_{3}$ &$0.4$& $0.9$& $0.4$\\
$y_{4}$ & $0.3$ &$0.6$& $0.8$ & $y_{4}$ &$0.8$& $0.5$& $0.8$&$y_{4}$ &$0.3$& $0.7$& $0.8$\\
\hline
\end{tabular}
\end{table}
\end{example}

Dynamic fuzzy relation systems can be compressed as the original
information system with high computational complexity. By using the
proposed algorithm, we compress them with low computational
complexity. Obviously, the proposed algorithm provides an effective
approach to compressing dynamic FRIS.

\subsection{Variations of fuzzy relation values}

We denote the revised $R$ as $R^{\star}$ when changing the relation
values between $x\in U_{1}$ and other objects in
$(U_{1},\mathscr{R})$. Subsequently, we compress
$(U_{1},\mathscr{R}^{\star}_{1})$, where
$\mathscr{R}^{\star}=\{R^{\star}\}\cup \mathscr{R}/R$. There
are four cases to be considered.

(1): $|[x]_{R}|=1$ and $|[x]_{R^{\star}}|=1$

In this case, the compression of the dynamic fuzzy relation
information system is the same as that of the original fuzzy
relation information system.

(2): $|[x]_{R}|=1$ and $|[x]_{R^{\star}}|>1$

In other words, the change of relation values classifies the object
$x$ into another class. If we have $y\in [x]_{R^{\star}}$ and $y\neq
x$, then $x$ can be mapped into the same image as $y$ under the
condition of the consistent function.

(3): $|[x]_{R}|>1$ and $|[x]_{R^{\star}}|=1$

That is to say, the variation of relation values constructs a new
class only containing the object $x$, and $x$ is mapped into a new
image.

(4): $|[x]_{R}|>1$ and $|[x]_{R^{\star}}|>1$

The change of relation values classifies $x$ into other class. If
$y\in [x]_{R^{\star}}$ and $y\neq x$, then $x$ is mapped into the
same image as $y$.

\section{Conclusions}

Information system homomorphism is an effective approach to
attribute reduction. In this paper, we have investigated more
properties of consistent functions and proposed an incremental
algorithm for constructing homomorphisms between fuzzy relation
information systems, which can be applied to compress dynamic fuzzy
relation information systems. After that, by using the precious
compression of the original FRIS we have
compressed dynamic FRIS. The
experimental results have illustrated that the proposed algorithm
had provided an efficient approach to compressing fuzzy relation
information systems.

In the future, there are many questions worthy of consideration. For
example, we will propose more effective algorithms for constructing
homomorphisms between information systems and apply them to compress
information systems. Furthermore, we will focus on the development
of effective approaches for attribute reduction and other tasks of
dynamic information systems.

\section*{ Acknowledgments}

We would like to thank the anonymous reviewers very much for their
professional comments and valuable suggestions. This work is
supported by the National Natural Science Foundation of China (NO.
11071061) and the National Basic Research Program of China
(2011CB311808).

\end{document}